\documentclass{ws-rv975x65}
\usepackage{subfigure}     
\usepackage{ws-rv-van}     
\usepackage{fix-cm}
\usepackage{amsmath}
\usepackage{amssymb}
\usepackage{amsfonts}
\usepackage{bm}
\usepackage[colorlinks=true, citecolor=blue,linkcolor=blue]{hyperref}

\def\sect#1{Sec.~\ref{#1}}
\def\sss{\scriptscriptstyle}
\def\UV{\scriptscriptstyle UV}
\def\st{\begin{equation}}
\def\stp{\end{equation}}
\def\Fig#1{fig.~\ref{#1}}

\def\half{\frac 12}

\newcommand{\Tr}[1]{\textrm{Tr}\left[#1\right]}
\def \tr {\mathrm{Tr}}

\def \bx {\mathbf{x}}

\def \nc {N_c}
\def \ca {C_A}
\def \nf {n_f}

\newcommand{\Tint}[1]{{\hbox{$\sum$}\!\!\!\!\!\!\!\int\,}_{\!\!\!\!\raise-0.9ex\hbox{$\scriptstyle{#1}$}}}
\def\onetwo{{1\lra2}}
\def\twotwo{{2\lra2}}

\def \cc {\mathcal{C}}
\def\siml{{\ \lower-1.2pt\vbox{\hbox{\rlap{$<$}\lower6pt\vbox{\hbox{$\sim$}}}}\ }}
\def\simg{{\ \lower-1.2pt\vbox{\hbox{\rlap{$>$}\lower6pt\vbox{\hbox{$\sim$}}}}\ }}

\def \bx {\mathbf{x}}

\def \crr {C_R}

\def \bfnabla {\boldsymbol{\nabla}}

\def \cc {\mathcal{C}}

\def \als {\alpha_{\mathrm{s}}}

\def \m2   {\mu^{2 \epsilon}}

\newcommand{\order}[1]{\mathcal{O}\left(#1\right)}

\def\siml{{\ \lower-1.2pt\vbox{\hbox{\rlap{$<$}\lower6pt\vbox{\hbox{$\sim$}}}}\ }}
\def\simg{{\ \lower-1.2pt\vbox{\hbox{\rlap{$>$}\lower6pt\vbox{\hbox{$\sim$}}}}\ }}
\def\rrangle{\right\rangle}
\def\llangle{\left\langle}

\def\nn {\nonumber}

\def\pp {p_\perp}

\def\qp {q_\perp}
\def\bpp {\mathbf{p}_\perp}
\def\bqp {\mathbf{q}_\perp}
\def\kp {k_\perp}
\def\bkp {\mathbf{k}_\perp}
\def\ql {\hat{q}_L}
\def\mmf {m_\infty^2}
\def\mmg {m_\infty^2}

\def\calr {\mathcal{C}_R}
\def\cala {\mathcal{C}_A}
\def\lra{\leftrightarrow}

\def\Eq#1{Eq.~(\ref{#1})}

\def \bh {\mathbf{h}}
\def \bff {\mathbf{F}}
\def \b {\mathbf{b}}
\def \f {\mathbf{f}}
\def \OO {\mathcal{O}}

\def \qm {q^-}
\def \qll {q^+}

\def \kl {k^+}

\def \cc{{\mathcal C}}

\def\h{{\bm h}}

\def\x{{\bm x}}
\def\p{{\bm p}}
\def\q{{\bm q}}
\def\k{{\bm k}}
\def\v{{\bm v}}

\def\D{{\bm D}}
\def\f{{\bm f}}

\def\nbe {n_\mathrm{B}}

\def\qhat {\hat{q}}

\def\NLO{{\mathrm{NLO}}}
\def\coll{{\mathrm{coll}}}

\def\md {m_D}
\def\mE {m_{\sss E}}

\def\xp {x_\perp}
\def\bxp {\mathbf{x}_\perp}

\makeindex
\begin{document}

\chapter[Parton energy loss and pT broadening at NLO in  high temperature QCD]
{Parton energy loss and momentum 
broadening at NLO in high temperature QCD plasmas}\label{ra_ch1}

\author[J. Ghiglieri and D. Teaney]{Jacopo Ghiglieri}

\address{Institute for Theoretical Physics, Albert Einstein
  Center,\\ University of Bern, Sidlerstrasse 5, 3012 Bern, Switzerland\\
  jacopo.ghiglieri@itp.unibe.ch}

  \author[J. Ghiglieri and D. Teaney]{Derek Teaney}
\address{Department of Physics and Astronomy, Stony Brook University,\\
Stony Brook, New York 11794-3800, United States\\
  derek.teaney@stonybrook.edu}

  \begin{abstract}
We present an overview of a perturbative-kinetic approach
to jet propagation, energy loss, and momentum broadening in a 
high temperature quark-gluon plasma.
The leading-order kinetic equations describe the interactions
between  energetic jet-particles and a non-abelian plasma,  consisting
of on-shell thermal excitations and
 soft gluonic
fields. These interactions include
$\twotwo$ scatterings, collinear bremsstrahlung, and drag and momentum diffusion.
We show how the contribution from the soft gluonic fields
can be factorized into a set of Wilson line correlators 
on the light cone. We review
recent field-theoretical developments, 
rooted
in the causal properties of these correlators, which 
simplify the calculation of the appropriate Wilson lines
in thermal field theory.
With these simplifications lattice measurements of transverse momentum
broadening have become possible, and the kinetic equations
describing  parton transport have been extended to next-to-leading
order in the coupling $g$.
\end{abstract}

\body

\section{Introduction}
\label{intro}
The suppression of highly energetic jets (or jet quenching) is
one of the most striking findings of the experimental program of
heavy-ion collisions~\cite{Roland:2014jsa,d'Enterria:2009am,Majumder:2010qh,Mehtar-Tani:2013pia,Baier:2000mf}.
A comprehensive review\cite{xinnian} can be found in this volume,
which also contains another contribution
reviewing a specific approach in greater detail~\cite{Blaizot:2015jea}.

In this review, we concentrate on a weakly-coupled
kinetic approach describing the
propagation of high momentum jet-like particles through a Quark-Gluon Pasma (QGP).
A detailed perturbative description of the QGP and jet-quenching is available
when the temperature is  high $T\gg \Lambda_{QCD}$, and
the momentum of the jet-particles is much larger than the
temperature, $p \gg T$.

 The kinetic picture of the high temperature QGP, which has emerged
through a combination of physical intuition, direct diagrammatic analysis, and
a gradient expansion, is characterized by hard particles and
random classical non-abelian fields~\cite{Braaten:1989mz,Braaten:1991gm,Blaizot:2001nr,Arnold:2002zm}. 
We will describe 
a Boltzmann equation for the propagation of jet-like particles 
interacting with the hard particles and random fields which 
comprise this idealized weakly coupled plasma.
Here the jet-particles
have momentum $p\gg T$, while the hard particles have momentum  $\sim T$,  and
the random classical fields are soft with momenta of order $gT$, $p \gg T \gg gT$, where $g$ is the strong coupling constant. 
For simplicity, we will limit the discussion to pure gauge theory.
As discussed in  Section~\ref{chap_classical},
to leading order in the coupling constant there are three
processes 
relevant to  the transport of jet-particles in the high 
temperature plasma:
(i)  $\twotwo$ elastic scatterings
with hard particles,  (ii) collinear radiation, and  (iii) drag and momentum diffusion  driven by the soft classical background.

The interactions between the jet-particles and the classical non-abelian fields, 
requires a resummation scheme known as the Hard Thermal Loop (HTL) effective
theory \cite{Braaten:1989mz,Braaten:1991gm},  which is the QCD analog
of the Vlasov equations \cite{Blaizot:2001nr}.
This effective theory is necessary to compute the 
drag coefficients and collinear bremsstrahlung
rates which describe the propagation of jets.
 The computational complexity of 
the Vlaslov equations would, at first sight, make any extension beyond leading order in the coupling 
extremely challenging. However, 
Hard Thermal Loops correlators (and statistical correlators more generally)
simplify greatly when evaluated at lightlike
separations~\cite{CaronHuot:2008ni}. It is precisely such light-like correlators
which must be evaluated to determine the
drag, diffusion, and collinear bremmstrahlung rates of jet-particles
propagating in plasma.

These simplifications are a consequence of the following
physical picture:  Since the hard and jet-particles are propagating 
almost exactly along the light cone, they are probing an essentially
undisturbed plasma, at least as far as the soft classical background
is concerned.  Informally, we say that the soft classical background 
``can't keep up'' with the hard or jet-particle traversing the plasma.
Thus, the soft correlations that are probed by 
the hard and jet-particles are statistical in nature rather 
than  dynamical. 

In Sec.~\ref{chap_sumrule}
we will illustrate the basics of Hard Thermal Loops and then proceed to
introduce the field-theoretical arguments for the light-like simplifications,
originally due to Caron-Huot\cite{CaronHuot:2008ni}. We then use
these simplifications at leading order to determine the transverse momentum diffusion and drag coefficients for high momentum probes.

An important consequence of these lightlike simplifications and their physical origin (see above)
 is that lattice  techniques
can be used to compute several lightlike correlators,
which arise in jet-quenching physics. Due to the
Minkowskian nature of the problem and the large energies at play, a direct
ab-initio lattice calculation of jet propagation in medium is not feasible.
However, as we show in detail, the soft contribution
to quantities such as $\hat{q}$ (the jet quenching parameter) becomes
Euclidean and three-dimensional, which makes it amenable to
lattice measurements. In Sec.~\ref{chap_lattice} we  illustrate
the basic principles of the dimensionally reduced (three-dimensional)
effective theory, and review the first lattice calculation\cite{Panero:2013pla}
of the soft contribution to $\qhat$.

Of course, the special properties of HTLs on the lightcone also simplify
perturbative calculations. Indeed, Caron-Huot employed these results to 
compute $\qhat$ and the transverse scattering kernel to 
NLO\cite{CaronHuot:2008ni} in the strong coupling constant $g$. Building on this and related developments,
the photon and low-mass dilepton emission rates were also computed to NLO~\cite{Ghiglieri:2013gia,Ghiglieri:2014kma}.
These electromagnetic rates
at NLO correct the strictly collinear approximations made at leading order.
Finally, this review anticipates an upcoming publication by the current authors where the LO Boltzmann equation for jet-transport described above is extended to NLO~\cite{jetwip}.
In Sect.~\ref{sec_nlo} we sketch the Boltzmann equation to 
this order, which  includes the drag coefficient at NLO and semi-collinear emission rates describing bremsstrhalung at  wider angles.  
Many of the parameters of the effective kinetic theory, such 
as drag coefficients,  thermal masses, and the transverse collision kernel, can be computed
using the three-dimensional lattice described above, providing a 
tantalizing semi-perturbative description of energy loss. 
These results and future directions are summarized in Sec.~\ref{sec_concl}.

\section{The kinetic picture at leading order}
\label{chap_classical}

In this section we will summarize the kinetics of high temperature weakly
coupled non-abelian plasmas.   As we mentioned in the introduction,
this kinetic picture of the QGP 
is characterized by hard particles and
random classical non-abelian fields~\cite{Braaten:1989mz,Braaten:1991gm,Blaizot:2001nr,Arnold:2002zm}. 
First we will
describe the interactions between these constituents qualitatively, and
subsequently give a more quantitative description.
We are particularly interested in formulating a Boltzmann equation 
for the transport of high momentum gluons $p\gg T$ traversing  
the QGP, $i.e.$ a Boltzmann equation for jet-particles 
interacting with the hard particles and random classical fields which 
comprise the high temperature plasma. 
The phase space
distribution of the jet-particles is notated with $f(t,\x, \p)$ (or $f_\p$ when clear from context) while
the phase space distribution  of the equilibrium hard particles
is notated with $n(t, \x, \p)$ (or $n_\p$).

The interactions between the particles and the classical 
fields of the QGP 
are characterized by three processes: 
\begin{enumerate}
   \item First,
there are $2\leftrightarrow 2$ collisions between the 
hard and the jet-particles.  These scattering events are strongly localized
in space-time.

   \item Second, the random classical fields in the plasma induce
      drag  and
      momentum diffusion and change the momenta of the jet-particles in
      characteristic ways. The drag and diffusion
      coefficients are calculated by examining the response
      of the classical fields to the non-equilibrium jet-particle.

   \item Finally, the momentum diffusion produced by the random classical
      field induces collinear bremsstrahlung, causing the jet-particles
      to split collinearly.
\end{enumerate}
At leading order in the coupling $g$, the 
Boltzmann equation for the jet-particles takes the form
\st
\label{loboltzmann}
\left(\frac{\partial}{\partial t}  + \v_\p \cdot \frac{\partial}{\partial \x}\right) f(t,\x, \p) = C_{2 \leftrightarrow 2}[\mu] + C_{\rm diff}[\mu] +  C_{\rm coll} \, , 
\stp
where the three rates on the right-hand-side reflect these three processes.
We have anticipated that the $\twotwo$ and diffusion rates 
depend on a separation scale $\mu$, but the dependence on $\mu$ cancels
when both processes are included.
In the remainder of this section we will discuss each of these rates
in greater detail.

The hard particles are approximately on-shell and carry the majority
of the energy and momentum of the plasma. For this 
reason these hard constituents are the most important for
the thermodynamics of the QGP, and the soft fields
are only important insofar as they influence the kinetics of the 
jet and hard particles. 
The typical momentum of these excitations is of order the temperature,  but the virtuality is
of order $\sim g^2 T^2$. Specifically, for a hard particle moving in the 
positive $z$ direction, the
scaling of momentum in lightcone coordinates is\footnote{ We use
   a notation where $p^+ = \frac{p^0 + p^z}{2}$,  $p^- = p^0 - p_z$, four-vectors
   are denoted by uppercase letters and
   the metric is the ``mostly-plus'' one, 
   so that $P^2= - 2 p^+ p^- + p_\perp^2$. The integration measure is
   \st
   \int \frac{d^4P}{(2\pi)^4} = \int \frac{dp_{+} dp_{-} d^2\pp }{(2\pi)^4 }.
   \stp
}
\begin{align}
\label{onshell}
   P^{\mu} =& \left(p^+, p^-, \p_{\perp} \right) \, ,  \\
   =& \left(T, g^2 T,  gT \right).
\end{align}
Squaring this momentum  shows that the virtuality of 
these on shell constituents is of order $P^2  \sim  g^2 T^2$.
To estimate the contribution of these on-shell constituents to 
the plasma energy density we use the equilibrium distribution
function, $n_{\p} =  1/(e^{E_p/T} - 1)$, and write
\st
e(t,\x) =  2d_A \int_{\sim g T}^{\infty} \frac{d^3p}{(2\pi)^3} E_\p n_{\p} \, .
\stp
First, we note that the integral is cut  off when the three-momentum 
of the constituents is soft, $p\sim gT$. 
At this point 
the power counting implicit in \Eq{onshell} no longer applies.
Taking  the free gas  dispersion relation $\epsilon_\p = p$, we 
see that the contribution of the soft modes ($p\siml g T$) is of order $g^3 T^4$,  while the leading contribution is of order $T^4$. 

The first correction 
to the free gas result for thermodynamic properties the QGP 
can be  found by determining the correction to the dispersion curve
for the on-shell particles.
Specifically, the dispersion curve
(the relation between $E_\p$ and $\p$)
is found by evaluating the one-loop self energy  for 
an approximately on shell gluon.
The self-energy diagram involves internal lines of momentum
\st
Q^{\mu} = (q^+, q^-, q_{\perp})  \sim (T, T, T) \, ,
\stp
and are thus highly virtual compared  to the on-shell particles.
The effect of these virtual modes (which are integrated out in
any kinetic description) is to change the dispersion relation
of the  on-shell modes
\st
  E_\p = p + \frac{m_\infty^2}{2p} + \ldots \, ,
\stp
where the ellipses indicate higher order corrections.  Here we have
defined the leading order gluon asymptotic mass  in pure gauge 
theory at high  temperature~\cite{Kalashnikov:1979cy,Weldon:1982aq,Blaizot:2005wr,CaronHuot:2008uw}
\st
\label{eq:asymptoticm}
m_{\infty}^2 = 2 g^2 C_{A} \int \frac{d^3p}{(2\pi)^3} \frac{n_{B}}{p} =  \frac{1}{6} g^2 C_A T^2 \,. 
\stp
As discussed in Sec.~\ref{sec_nlo}, the asymptotic mass can 
be expressed as a correlator of field strengths along the light cone,
which is closely related to an analogous correlator for $\hat q$, see \Eq{asymmass}.

\begin{figure}
   \begin{center}
   \includegraphics[width=0.4\textwidth]{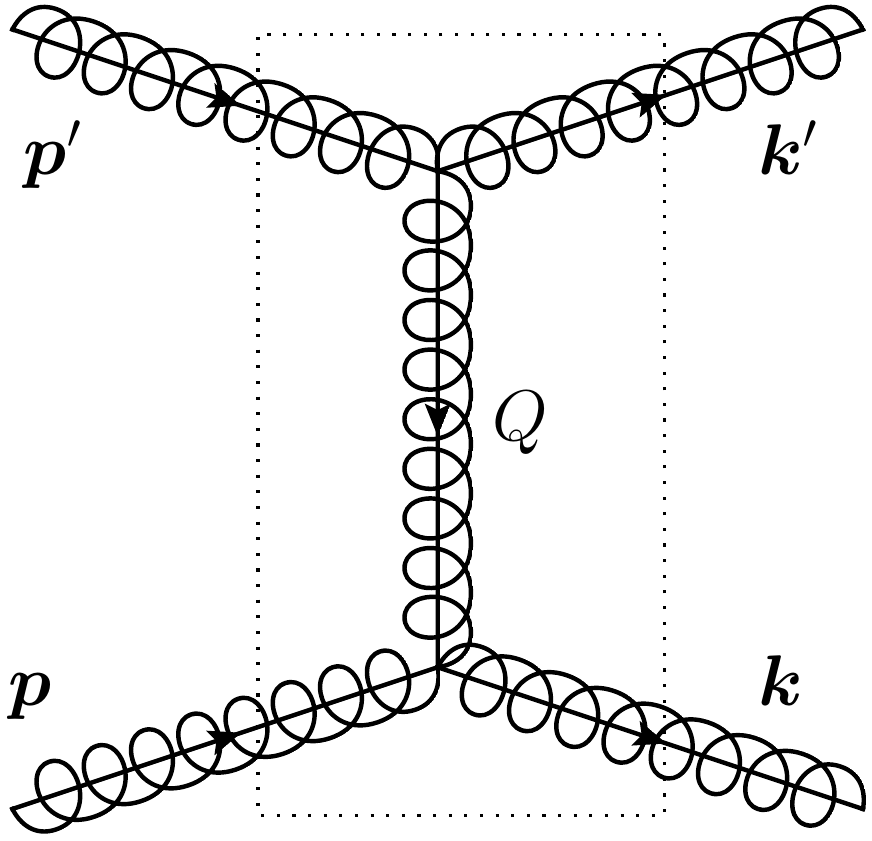}
\end{center}
   \caption{ Hard $2\leftrightarrow 2$ collision contributing 
      the collision rate $C_{2\leftrightarrow 2}[\mu]$. 
      Only hard lines which enter or exit the boxed region are included
      in an effective Boltzmann description.
   \label{hard2to2} }
\end{figure}

These hard modes with $p\sim T$ occasionally collide with the jet partons,
transferring momenta of order the temperature.
At leading order 
these collisional processes are $2\leftrightarrow 2$ processes, 
which can be localized in space-time to within a distance of $\sim 1/T$.
A kinetic equation
for the phase space distribution function of the jet partons including such
scattering processes takes the form 
\st
\left(\frac{\partial}{\partial t}  + \v_\p \cdot \frac{\partial}{\partial \x}\right) f_\p(t,\x) = C_{2 \leftrightarrow 2}[\mu] + \ldots,
\stp
where the ellipses will be described below.  The velocity  in
this equation is the group velocity, $\v_\p  = \partial E_\p/\partial \p$.
The collision
rate between the hard modes and the jet particles takes the 
form\footnote{We use a familiar shorthand 
	notation, $\int_{\p\k\k'} \equiv \int \frac{d^3p}{(2\pi)^3} \frac{d^3k}{(2\pi)^3} \frac{d^3k'}{(2\pi)^3}$, and we are neglecting
   the thermal mass when writing \Eq{eq:collision22exp}. The 
   leading factor of $1/(4d_A)$ reflects an average over spins
   and colors of the external gluon and a symmetry factor $1/2$
   for the sum over final states.
}
\begin{multline}
   -C_\twotwo[\mu]
   = \frac{1}{4d_A}
\int_{\p'\k\k'}
   \frac{|\mathcal M|^2}{(2 p) (2 p') (2 k) (2k') }  (2\pi)^4 \delta^4(P+P' - K - K') \\
\times \Big[
   f_\p  n_{\p'} (1{+}n_{\k}) (1{+}n_{\k'})
   {-}
   f_{\k} n_{\k'} (1{+}n_{\p'})
{-}
n_{\k} f_{\k'} (1{+}n_{\p'})
\Big] \,,
\label{eq:collision22exp}
\end{multline}
where in the second line we have dropped terms that are exponentially
suppressed for $p\gg T$.\footnote{We consider $\exp(-p/T)\ll1$. In the remainder
of this review we sometimes treat $T/P$ as an explicit expansion parameter for 
illustration purposes. The detailed exposition and the calculations in\cite{jetwip} 
are performed without treating $T/P$ as an expansion parameter.} The kinematics of the collisional process with momentum 
exchange of order $\sim T$ is shown in \Fig{hard2to2}.
The squared matrix element (summed over initial and final  state colors and helicities) may be obtained from vacuum perturbation 
theory:
\st
|\mathcal M|^2 =  16 d_{A} g^4 C_{A}^2  \left(3 - \frac{su}{t^2} -\frac{st}{u^2} - \frac{tu}{s^2} \right) \, .
\stp

The $2 \leftrightarrow 2$ scattering rate is divergent in the infrared, 
and therefore we have notated the dependence on the 
separation scale $\mu$ in \Eq{eq:collision22exp}.
There are many ways this divergence can be regulated. At leading
and next-to-leading we find it convenient~\cite{jetwip} to
simply cutoff the transverse momentum exchange at small $q_{\perp}$, $q_\perp > \mu$.
It is not difficult to extract the logarithmic dependence on $\mu$ for $\mu \ll T$.
Indeed, let us consider for illustration a leading-log approximation to $C_{\twotwo}[\mu]$:
we expand the distribution
function and matrix elements  to second order in the exchange momentum $Q$
and arrive at a Fokker-Planck equation\cite{Svetitsky:1987gq,Arnold:2000dr,Hong:2010at} for $f_\p$ 
\st
\label{eq:twotwoexpand}
   C_{\twotwo}[\mu] =  \hat e_{\scriptscriptstyle UV}(\mu) \, v^i \frac{\partial f_\p }{\partial p^i }
   +   \half  \hat q^{ij}_{\scriptscriptstyle UV}(\mu) \frac{\partial^2 f_\p}{\partial p^i \partial p^j } + \mathcal O\left(\frac{T}{p}\right) + \mbox{$\mu$-independent}\, , 
\stp
In writing this equation we have dropped terms 
suppressed by $T/p$.
Here $\hat v$ is a unit vector in the direction of $\p$, and the diffusion tensor $q^{ij}_{\UV}(\mu)$ controls the 
longitudinal
and transverse momentum 
diffusion,
\st
\hat q^{ij}_{\UV} (\mu) \equiv \hat q_{L,\UV}(\mu) \hat v^i \hat v^j  + \half \hat q_{\UV}(\mu) (\delta^{ij} -  \hat v^i \hat v^j) \, .
\stp
The values of these coefficients are found from  the expansion
of \Eq{eq:collision22exp}, and for pure gauge are at leading log
\begin{align}
   \label{uvqhat}
   \hat q_{\UV}(\mu) =& g^2 C_A T \frac{m_{D}^2}{2\pi} \log \left(\frac{ T}{\mu} \right) \, ,\\
   \hat q_{L,\UV}(\mu) =& g^2 C_A T \frac{m_{\infty}^2}{2\pi} \log \left(\frac{ T}{\mu} \right) \, .
\end{align}
Here the Debye mass is given by the integral over distribution functions
\st
   \label{eq:md}
   m_{D}^2 = 2g^2 C_A \int \frac{d^3p}{(2\pi)^3} \frac{n_p (1 + n_p) } {T} = \frac{1}{3} g^2 C_A T^2 \, ,
\stp
and the asymptotic mass is given by a similar integral in \Eq{eq:asymptoticm}.
At this point the interpretation of these thermodynamic integrals 
as the 
Debye and asymptotic masses is premature. This interpretation 
will be clear from \sect{chap_sumrule},  which 
explains physically why the transverse diffusion coefficient involves the Debye
mass, while the longitudinal diffusion involves the asymptotic mass.

The fluctuation-dissipation theorem
relates the coefficient of $\hat q_{L,\UV}(\mu)$ 
to $\hat e_{\UV}(\mu)$
\st
\label{einsteinrel}
q_{L,\UV}(\mu) = 2 T \hat e_{\UV}(\mu)   + \mathcal O\left(\frac{T}{p} \right) \, .
\stp
This Einstein relation relation is necessary in order 
for $f_\p$ to reach its equilibrium form $e^{-p/T}$ under the Fokker-Planck
evolution in \Eq{eq:twotwoexpand}.  Since the equilibrium distribution is a
stationary solution to the full $2\leftrightarrow 2$ Boltzmann equation
(\Eq{eq:collision22exp}) this relation between the drag and the longitudinal
diffusion coefficient automatically arises from a direct expansion of the Boltzmann equation.
\cite{Arnold:1999uza,Arnold:1999va,jetwip}

As we will see, the cutoff dependence 
in the $2\leftrightarrow 2$ contribution to the Boltzmann equation
cancels when the interactions with the soft background fields are
included. The description of these soft modes is our next task.
As discussed above, the virtuality of the hard modes is of order $P^2 \sim (g T)^2$
with the  momentum scaling given in \Eq{onshell}. The
soft gluonic excitations, on the other hand, have a momentum scaling
\st
\label{soft}
P^{\mu} = (g T, g T,  gT) \, , 
\stp
but also have virtuality of order $P^2 \sim (g T)^2$.
Since these modes are soft, they are highly occupied due to the Bose-Einstein distribution function
\st
\label{classicbose}
n_B(p^0) \simeq \frac{T}{p^0} \sim  \frac{1}{g} \, .
\stp
Thus, at leading order these soft modes can be treated
as a classical gluonic field interacting with the  
hard on-shell modes.

It is important  to emphasize that the energy in these
soft modes constitutes a small fraction of the total energy density. Nevertheless,
they significantly influence the kinetics of the on-shell particles.
Specifically, these soft modes collide frequently with
the hard and jet particles and exchange soft momenta of order $gT$.
It seems intuitive that the 
effect of these soft scatterings can be incorporated into a Fokker-Planck
equation describing the drag and momentum diffusion of the jet (and hard) particles
\st
\left(\partial_t + v_\p \cdot \frac{\partial}{\partial \x} \right) f_{\p} = C_{\rm diff}[\mu]  + \ldots \, , 
\stp
The collision (or Fokker-Planck) operator can be written as
\st
\label{fp}
C_{\rm diff}[\mu]  = \hat e(\mu) \, v^i \frac{\partial f }{\partial p^i }
+   \half  \hat q^{ij}(\mu) \frac{\partial^2 f}{\partial p^i \partial p^j }+\order{\frac TP},
\stp
where the momentum diffusion parallel and perpendicular to the direction of 
motion $\p$ are quantified by $\hat q^{ij}(\mu)$ 
and the fluctuation-dissipation theorem is satisfied as described above\footnote{In the remainder of this
	review $\hat q$ describes only the 
contribution of soft $g T$ to transverse momentum broadening, and perhaps 
should be written as $\hat q_{\rm soft}$. In most of the literature on
jet energy loss  (see for example ~\cite{Arnold:2008vd}) $\hat q$ refers to the sum of the soft and hard  contribution,
with a logarithmic sensitivity on the jet energy $p$.}.
Terms of higher order in $T/p$ are discussed in detail elsewhere\cite{jetwip}.

We have emphasized that the parameters of the Fokker-Planck 
evolution depend on the scale $\mu$ separating the scattering from
soft modes,  and the scattering from hard $2\leftrightarrow2$ 
collisions.
As the separation scale is changed, the parameters of the Fokker-Planck 
equation change and the drag and diffusion rates change accordingly. Indeed,
\sect{chap_sumrule} shows that at leading order the drag and 
diffusion coefficients are
\begin{subequations}
   \label{eq:qlosummary}
\begin{align}
   \hat q(\mu) =& g^2 C_A T \frac{m_{D}^2}{2\pi} \log \left(\frac{ \mu}{m_D} \right) \, ,  \\
   \hat q_{L}(\mu) =& g^2 C_A T \frac{m_{\infty}^2}{2\pi} \log \left(\frac{ \mu}{m_{\infty}} \right) \, .
\end{align}
   \end{subequations}
Thus, a change in these coefficients due to a change in $\mu$ is compensated by a corresponding change in the hard $2\leftrightarrow2$ collision operator in \Eq{eq:twotwoexpand}.  

As is well known the Fokker-Planck evolution specified by \Eq{fp} is equivalent to a Langevin process~\cite{kittel2004elementary},  where each particle in
the phase space follows
an equation of motion specified by a drag and a stochastic force $\xi^{i}$, whose variance is specified by the diffusion coefficients
\st
\label{langevin}
\frac{dp^i}{dt}  = - \hat e(\mu) v^i +  \xi^i \,,    \qquad  \llangle \xi^i(t) \xi^j(t') \rrangle = \hat q^{ij}(\mu) \delta(t - t') \, . \    
\stp
The Langevin description is valid on timescales which are long compared
to the underlying correlation times between the microscopic forces
in the medium.  For a classical particle satisfying
Newton's Law, $dp^i/dt = \mathcal F^i$,  one would adjust 
the diffusion coefficient so that the integrated 
squared variance  between the microscopic and stochastic forces
agree
\st
\hat q^{ij} \mathcal T   \equiv \int dt \int dt' \, \llangle \xi^i(t) \xi^j(t') \rrangle  = \int dt \int dt' \,  \llangle \mathcal F^i(t) \mathcal F^j(t') \rrangle\, .
\stp
Here $\mathcal T$ is the total time,
and the forces are to be evaluated
along the trajectory of the particle, $i.e.$ for a particle
traveling at the speed of light along the $z$-axis the coordinates are $x^{\mu} =(x^+, x^-, \x_\perp)= (t, 0, {\bm 0}_\perp)$.
For a hard particle in representation $R$ interacting with a soft classical background QCD field, the forces are  $g T_R^a F_a^{i\mu}(x^+)v_{\mu} $ and are 
dressed with Wilson lines following the trajectory of the particle along the light cone from past infinity\footnote{Here and below we use a matrix notation 
$F^{\mu\nu} = F^{\mu\nu}_a T^a_R$. $\v^{\mu}$ is a lightlike four vector
$v^{\mu} = (1,{\bm v})$ in the direction of motion of the particle, which is
conventionally taken to be along the  $z$ axis. Thus, $F^{i\mu}v_{\mu} = F^{i-}$
and note that $F^{z-}= F^{+-}$. More explicitly, the Wilson lines are $U(x^+,0) = Pe^{-i \int_{0}^{x^+} dx^+ A^-}$. }
\st
\mathcal F^{i}(x^+) \equiv U^\dagger_R(x^+,-\infty) \, g F^{i\mu}(x^+)v_{\mu} \, U_R(x^+,-\infty) \, . 
\stp
The appropriate formulas for 
$\hat q$ and $q_L$ are then 
\begin{align}
\label{qT}
\hat q(\mu) =& \frac{1}{d_R} \int_{-\infty}^{\infty} dx^+ \; {\rm Tr} \llangle  \mathcal {\bm F}_\perp (x^+)  \mathcal {\bm F}_\perp (0) \rrangle \, , \\
\label{qL}
\hat q_L(\mu) =& \frac{1}{d_R} \int_{-\infty}^{\infty} dx^+ \; {\rm Tr} \llangle  \mathcal {\bm F}^z (x^+)  \mathcal {\bm F}^z (0) \rrangle \, ,
\end{align}
where the two transverse directions are summed over in \Eq{qT}, and
the trace averages over the colors of the incoming particle. 
For a classical background field these
correlators can also be written as%
\footnote{ To see this, take $dx^{+} = \Delta x^{+}$ small. 
   Then,  evaluate the following trace to first order in $\Delta x^+$ to see the structure of the adjoint Wilson line emerge
   \st
   \tr[\mathcal F^{z}(x^+) \mathcal F^{z}(0)] \simeq v_{\mu} F^{z \mu}_{a}(x^+) \, v_{\rho} F^{z\rho}_{b}(0) \, \tr[(1 + i g \Delta x^{+} A^{-})  \, T^{a} \, (1 - i  g\Delta x^+ A^{-}) \, T^{b} ]  \, .
   \stp
}
\begin{align}
\label{qT_versionb}
\hat q(\mu) =& \frac{g^2 C_R}{d_A} \int_{-\infty}^{\infty} dx^+  \llangle  v_{\mu} F^{\mu\nu}_{a} (x^+)  \, U_{A}^{ab}(x^+, 0) \, v_{\rho} F^{\rho}_{\phantom{\rho}\nu, b}(0)\rrangle \, , \\
\label{qL_versionb}
\hat q_L(\mu) =& \frac{g^2 C_R}{d_A} \int_{-\infty}^{\infty} dx^+ \llangle 
v_{\mu} F^{\mu+}_a(x^+) \, U_{A}^{ab}(x^+, 0)\, v_{\rho} F^{\rho+}_{b}(0) \rrangle \, ,
\end{align}
where $U_{A}^{ab}(x^+,0)$ is the adjoint Wilson line.

The evaluation of the  transport parameters $\hat q_{L}(\mu)$  and $\hat{q}(\mu)$ using these correlators  involves understanding 
how the hard particles interact with the classical gluon fields 
at leading and next-to-leading order. 
This had been carefully examined by Blaizot and Iancu who worked
in a background field gauge and systematically employed a gradient
expansion to determine the appropriate kinetic equations which 
are the non-abelian generalization of the Vlaslov equations~\cite{Blaizot:2001nr}.
We will write the non-abelian Vlaslov equations in the next
section and exhibit the appropriate HTL 
diagrammatic rules for 
evaluating these soft classical gluon correlators.  Then 
in \sect{chap_sumrule} we will show how these rules can be used to 
arrive at \Eq{eq:qlosummary}.

\begin{figure}
 \begin{center}
   \includegraphics[height=1.5in]{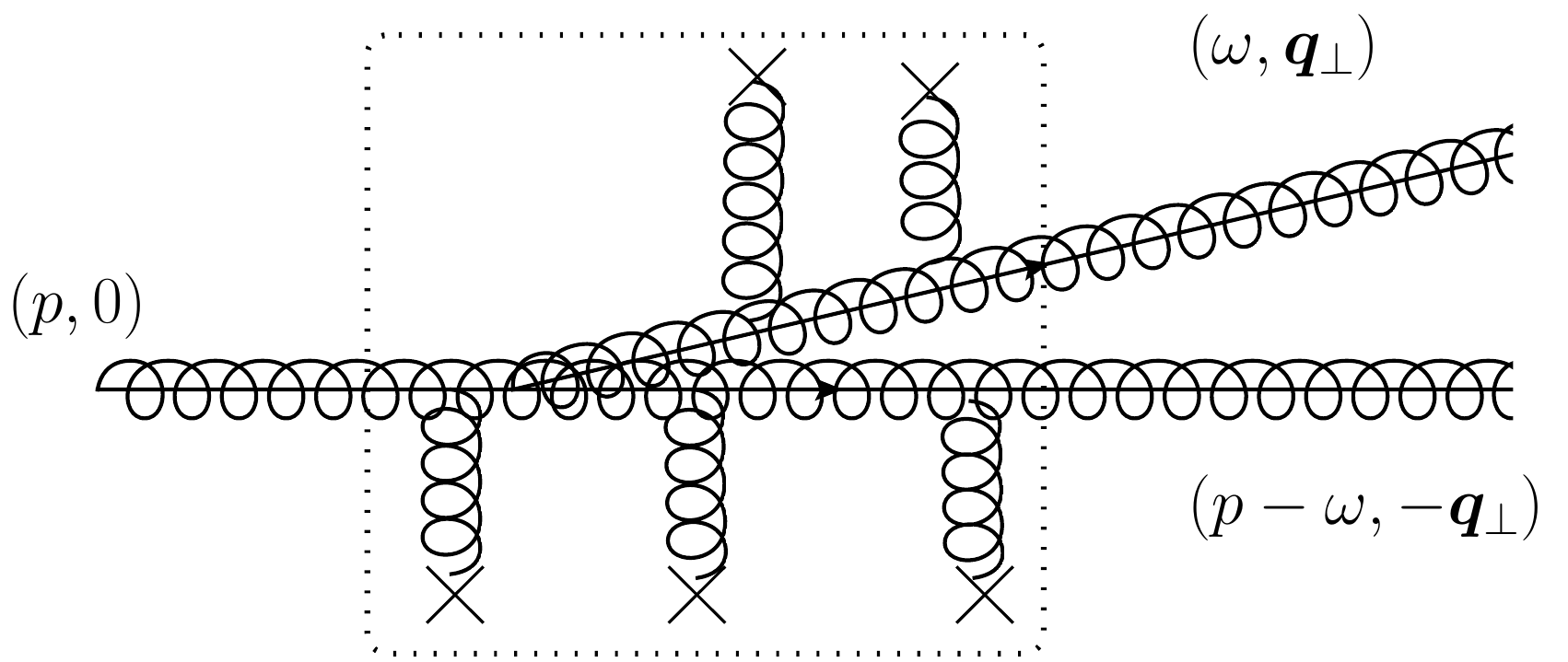}
\end{center}
   \caption{
      Schematic Feynman diagram contributing to the leading order 
      collinear bremsstrahlung rate. Hard gluon lines 
      are labeled by their three momentum $(p_z, \p_\perp)$. 
      The interactions with the random classical background
      bath are illustrated by the gluon lines with crosses. 
      Only hard lines which enter or exit the boxed region are included
      in an effective Boltzmann description.
      \label{locoll}
   }
\end{figure}

The random classical gluonic fields cause 
the hard and jet particles to diffuse in momentum.
This random walk  induces collinear bremsstrahlung in medium.
We will follow the AMY formalism for computing the collinear
bremsstrahlung rate~\cite{Arnold:2002zm}, and refer to ref.~\cite{Arnold:2008iy} which compares
the AMY formalism to other approaches, which are described in the reviews~\cite{Baier:2000mf,Majumder:2010qh}.
Collinear bremsstrahlung
can be included into the Boltzmann equation  and takes the
form of a local rate $C_{\rm coll}$ for a hard (approximately massless)
gluon to branch into two hard gluons moving in approximately the the same direction as their parent~\cite{Baier:2000sb,Arnold:2001ba,Arnold:2002ja}
\st
\left(\partial_t  + v_\p \cdot \partial_\x \right) f_\p = C_{\rm coll}  + \ldots \, .
\stp
To describe the gluon bremsstrahlung rate we refer to \Fig{locoll} 
which schematically illustrates the collinear radiation induced by the soft
scatterings. 
The kinematics are shown  for a gluon with momentum $p = |\p|$ in 
the longitudinal direction to branch into 
 two gluons moving 
in  approximately the same direction with longitudinal momentum $\omega$  
and $p-\omega$. 
Although the bremsstrahlung rate enters the Boltzmann equation as a local
rate, it must be understood that the emission process can only 
be localized to within a time scale set by the formation 
time of the radiation. The inverse formation time will be defined
as the energy difference between the initial and final states 
\st
(\tau_{\rm form})^{-1}  \equiv \delta E(h,p,\omega) =  (E_{\omega} + E_{p-\omega} ) - E_{p} \,.
\stp
Using the dispersion relation for the hard particles this reads
 \begin{align}
\label{defdeltaE}
\delta E(h,p,\omega) \simeq  
 \frac{h^2}{2p\omega(p-\omega)}+\frac{m^2_{\infty\,\omega}}{2\omega}
+\frac{m^2_{\infty\,p-\omega}}{2(p-\omega)}
-\frac{m^2_{\infty\,p}}{2p},
\end{align}
where $m_{\infty,p}^2$ is the asymptotic mass of the particle with momentum $p$, as summarized in Eq.~\eqref{eq:asymptoticm}.
We have further defined 
\st
 \h  \equiv  p \q_{\perp} \, .
\stp
As seen from the figure and described below, $\h/p$ is a transverse momentum vector 
which is conjugate to the (transverse) coordinate separation $\bxp$ between the initial and final states.

\begin{figure}
   \begin{center}
      \includegraphics[width=0.8\textwidth]{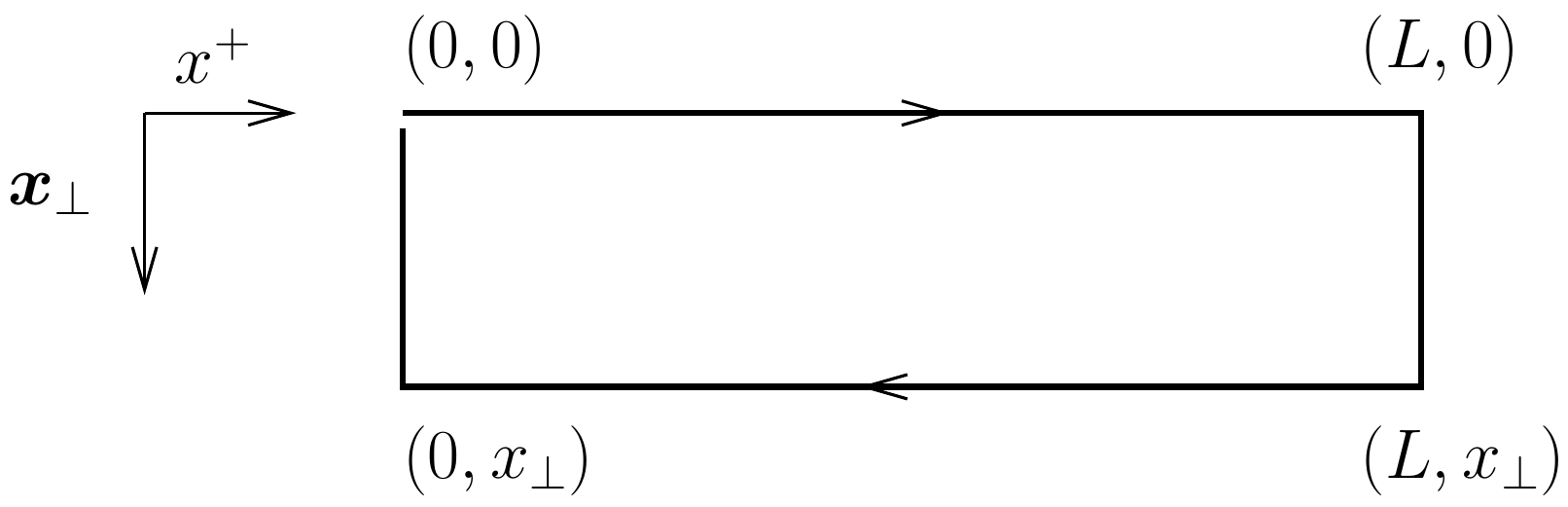}
      \caption{
       Wilson loop leading to $\hat q (\mu)$. Here the side 
       rails, $Pe^{-i\int_0^{L} dx^+ A^{-}(x^+,0)}$ and $Pe^{-i\int_L^{0} dx^+ A^-(x^+,x_\perp)}$, are separated by transverse coordinates $\bxp$ and 
       extend in $x^+ = (t + z)/2$ from $x^+ = 0$ up to $x^+=L$.
       $x^-$ is zero in this figure. The Wilson loop is
       traced over colors, $\frac{1}{d_R} \tr\ldots$, as indicated
       by the gap at the top left corner.
      \label{wilsonfig}
      }
\end{center}
\end{figure}

The bremsstrahlung rate
$C_{\rm coll}$ is determined by the 
rate of transverse momentum kicks (of magnitude $\qp$) which a
hard particle experiences traversing the soft classical fields:
\begin{equation}
\label{defcq}
\cc_R(\qp) \equiv 
\lim_{p\to\infty} (2\pi)^2 \frac{d\Gamma_R(\p,\p+\qp)}{d^2\qp} \, .
\end{equation}
Here $\p$ is the momentum of the hard particle, which 
is large ($\p \rightarrow \infty$) relative to the 
the typical momentum, $\sim gT$, of the background fields.  
The collision kernel $\cc_R$ can be expressed as a Wilson
loop in the $(x^+,x_\perp)$ plane evaluated in the classical background~\cite{CaronHuot:2008ni,D'Eramo:2010ak,Benzke:2012sz}, 
as sketched in Fig.~\ref{wilsonfig}.  To motivate
the appropriate Wilson loop
we note that the average  squared  momentum transfer per unit time ($i.e.$ $\hat q$)
is determined by $\cc_R(\qp)$
\begin{equation}
\label{qhat_LO}
\qhat(\mu)=\int^\mu\frac{d^2\qp}{(2\pi)^2} \qp^2 \cc_R(\qp) \, .
\end{equation}
As described above $\qhat$ can be expressed as a correlation of 
field strengths (see \Eq{qT}). Numerically at least, this correlation 
function should be understood as  
the limit of a Wilson loop (see \Fig{wilsonfig}) as the transverse 
separation approaches zero and the length approaches infinity
\st
\label{shortwloop}
\lim_{\xp \rightarrow 0 } \lim_{L\rightarrow \infty}  \llangle W_R(\xp,L) \rrangle = \exp\left( -  \frac{1}{4} \hat q(\mu) \xp^2 L  \right)
\stp
It is not difficult to show that this Wilson loop expression agrees with \Eq{qT} in the limit
of small separation.
Although it is not entirely obvious, it is certainly not surprising
that the Wilson loop is sufficient to completely determine 
$\cc_R(\q_\perp)$. Specifically, the impact parameter representation
of the collision kernel
\st
\label{defcb}
\cc_R(\xp)  = \int \frac{d^2q_\perp}{(2\pi)^2 } \left(1 - e^{i\q\cdot \bxp} \right)\, \cc_R(\q_\perp)\,, 
\stp
is determined by the asymptotic behavior of the same Wilson loop at
large $L$
\st
\label{cfromw}
 \lim_{L\rightarrow \infty} \llangle W_R(\xp,L)\rrangle  = \exp\left( -  \cc_R(\bxp) L  \right).
\stp
Expressions of this form have been used to compute $\hat q(\mu)$ at
leading and next-to-leading order.

We have described the collision kernel in detail, since it is a key parameter in determining the collinear bremsstrahlung rate. 
A derivation of this rate is beyond the 
scope of this review, and here we will simply quote 
the relevant formulas and describe the relevant physics.
The radiation rate  
for a gluon of momentum $p$ to radiate
a gluon of longitudinal momentum $\omega$ in approximately the same direction 
is\cite{Baier:2000mf,Arnold:2002ja}
\begin{eqnarray}
\nn\frac{d\Gamma(p,\omega)}{d\omega}\bigg\vert_\coll&=&\frac{g^2C_A}{16\pi p^7}
(1+ n(\omega))(1+ n(p-\omega))
\frac{1+x^4+(1-x)^4}{x^3(1-x)^3}\\
&&\times\int\frac{d^2h}{(2\pi)^2}2\bh\cdot\mathrm{Re}\,\bff(\bh,p,\omega),
\label{jmcoll}
\end{eqnarray}
where  $x\equiv \omega/p$
is the longitudinal momentum fraction of one of the outgoing gluons.
The leading factor $(1 + x^4 + (1-x)^4)/(x^3 (1-x)^3)$
records the collinear emission probability accompanying a scattering process.
Similarly $\bff(\bh,p,\omega)$ records the time 
evolution of the current over the formation time.
In momentum space this time evolution
results in an integral equation for $\bff(\bh,p,\omega)$
which resums the influence of multiple soft 
scatterings\footnote{Their effect on the emission process is usually 
called the Landau-Pomeranchuk-Migdal (LPM) effect.} on 
the emission process~\cite{Baier:2000mf,Arnold:2002ja}
\begin{eqnarray}
   \label{integraleq}
\nn2\bh&=&i\delta E(\bh,p,\omega)\bff(\bh)+\int\frac{d^2\kp}{(2\pi)^2}
\frac{\mathcal{C}_A(\kp)}{2}
\bigg\{[\bff(\bh)-\bff(\bh-\omega\bkp)]\\
&&+[
\bff(\bh)-\bff(\bh+p\bkp)]+[
\bff(\bh)-\bff(\bh-(p-\omega)\bkp)]\bigg\}.
\label{defimplfull}
\end{eqnarray}
Finally, the full contribution to the Boltzmann equation takes the form
of a gain term (reflecting the radiation from particles with momentum $p+\omega$)
and a loss term (reflecting the radiation from the particle with longitudinal momentum $p$)
\cite{Baier:2000sb,Jeon:2003gi}:
\begin{eqnarray}
    C_{\coll}&=& 
 \int_{-\infty}^{\infty} d\omega \, f_{p+\omega}\frac{d\Gamma(p+\omega,\omega)}{d\omega}\bigg\vert_\mathrm{coll} 
 - \theta(p - 2\omega) f_p \frac{d\Gamma(p,\omega)}{d\omega}\bigg\vert_\mathrm{coll}
\label{onetwocollision}
\end{eqnarray}
The region with $\omega < 0$ represents the collinear merging
of two gluons to make a final state gluon. We
also remark that \Eq{onetwocollision} is finite
in the infrared, as the power-law divergences
for $\omega\to 0$ and $\omega\to p$ cancel between the
gain and loss terms.

In practice, the integral equation for the bremsstrahlung rate (\Eq{integraleq}) is most easily solved by 
working in coordinate space, which also illustrates the role
of the coordinate space Wilson loop in shown in \Fig{wilsonfig}.
To facilitate this rewriting,
we recall $\h/p$ is conjugate
to $\x_\perp$, which we relabel as $\b$.
Dividing \Eq{integraleq} by $p$ and Fourier transforming with respect 
to $\h/p$ we arrive at a coordinate space representation of the 
integral equation:
\begin{eqnarray}
\nn-2i\bfnabla \delta^2(\b)&=&\frac{i}{2\,x\,(1-x)}\left(\mmg(1-x+x^2)
-\nabla_\b^2\right)\f(\b)\\
&&+
\left(\frac{\cc_A( x\b)}{2}
+\frac{\cc_A(\b)}{2}+\frac{\cc_A( (1-x)\b)}{2}\right)\f(\b),
\label{bspace}
\end{eqnarray}
where we have defined  
\begin{equation}
   \f(\b) = \int \frac{d^2h}{(2\pi)^2p^3} e^{i\frac{\b \cdot \bh}{p}}\bff(\bh)\,,
\end{equation}
and  note that the integral entering the rate \eqref{jmcoll} becomes
\begin{equation}
\label{b_want}
{\rm Re}\:\int \frac{d^2h}{(2\pi)^2 p^4} 2 \bh \cdot \bff(\bh)
= 2\, {\rm Im}(  \nabla_{b} \cdot \f(0))\,.
\end{equation}
This differential equation (\Eq{bspace}) can be solved to determine the bremsstrahlung rate.

The coordinate representation of the integral equation (\Eq{bspace}) naturally involves 
the Fourier-transformed  collision kernel $\cc_R(\b)$ stemming
from the Wilson line discussed above (see \Fig{wilsonfig} and \Eq{cfromw}).
Let us remark that the second line in Eq.~\eqref{bspace} contains a sum
of three two-body contributions, with the transverse distance
$b$ weighted by the ratio of two of the three energies $p$, $\omega$ and $p-\omega$.
Indeed, it is easy to show 
\cite{Baier:1994bd,Baier:1996kr,Zakharov:1996fv,Zakharov:1997uu,Arnold:2002ja}
that, at leading order, this is the only possibility. 
Beyond leading order, one could expect a more complicated
three-pole structure, with three Wilson lines on the light cone, properly linked 
by transverse lines at their endpoints. However, at least up to NLO 
the symmetries of the system
yield the simple planar dependence in the transverse plane reflected in \Eq{bspace}~\cite{CaronHuot:2008ni}.

To summarize, we have identified three processes at leading order
which are relevant to the transport of high momentum jet particles in a weakly
coupled QGP.  The first process is straightforward $\twotwo$ collisions. This rate depends on the separation scale $\mu$ between 
the $\sim T$ scale, characterized by hard particles, and the  $\sim gT$ scale,
characterized by classical fields. The second process is momentum diffusion, which describes the interactions between the classical fields
and the  jet-particles. This processes is described by a Fokker-Planck equation, whose parameters $\hat e(\mu)$ and $\hat q(\mu)$ also depend on the separation scale. However, the sum of 
the $\twotwo$ collisions and the Fokker-Planck evolution is independent of 
$\mu$. Finally, the third process is collinear bremsstrahlung, which is independent
of the separation scale to this order. The parameters entering the collinear bremsstrahlung rate are the asymptotic masses $m_{\infty}^2$ and the scattering kernel $\cc(q_\perp)$. Both
of these quantities can be expressed as (closely related) correlators of field strengths
along the light cone, Eqs.~\eqref{qT_versionb},\eqref{cfromw}, and\eqref{asymmass}.

Anticipating Sec.~\ref{sec_nlo}, at next-to-leading order we will find
corrections to the parameters of the Fokker-Planck equation, $\hat e(\mu)$ and
$\hat q(\mu)$, and corrections to the parameters of the collinear
bremsstrahlung rate, $m_{\infty}^2$ and $\cc(q_\perp)$.   As we
will see the NLO corrections to $\hat e(\mu)$ depend logarithmically on the separation scale $\mu$. However, at NLO
a correction to the collinear bremsstrahlung rate must also be included, which
accounts for not-so-collinear emissions.
This \emph{semi-collinear} emission rate also depends on the separation
scale $\mu$, and the complete 
NLO rate is independent of $\mu$.  Thus, the NLO correction to the collision  term takes the schematic form
\st
\delta C = \delta C_{\rm coll} + \delta C_{\rm diff}[\mu] + \delta C_{\rm semi-coll}[\mu] \, .
\stp

Before we describe these developments in detail, 
we will show how the computation of lightlike correlators in thermal field theory (such as $\hat e(\mu)$ and $\hat q(\mu)$) can be dramatically simplified 
using the  causal properties of these correlators. This is the task of the next section.

\section{Hard Thermal Loops on the lightcone: a modern perspective}
\label{chap_sumrule}
In the previous section we described a kinetic equation for jet-like particles based on 
$\twotwo$ collisions, diffusion and drag, and collinear bremsstrahlung.
The diffusion and drag coefficients are determined
by classical statistical correlators of the 
soft background field on the light cone, \Eq{qT} and \Eq{qL}. 
These correlation functions are to be evaluated  using the Hard Thermal Loop 
(HTL) effective theory, 
which describes the response of the coupled Maxwell-Boltzmann system.
 Similarly, the collision
kernel $\cc_{R}(\q_\perp)$ is also determined through \Eq{cfromw} by the interaction
between the classical particles and the the background field on the 
light cone.

Hard Thermal Loops involve a complex set of medium-modified propagators
and vertices.  Already at leading order the evaluation of the 
necessary correlators is seemingly complex. However, as we will describe
in this section the evaluation of \emph{lightcone} HTL correlators
turns out to be surprisingly simple.  For certain operators, such 
as those associated with transverse momentum diffusion \Eq{qT}, and 
soft corrections to the asymptotic mass,  a Euclidean
formalism can be used~\cite{CaronHuot:2008ni}. On the other hand,
other operators  such as those 
associated with longitudinal diffusion and drag are surprisingly 
insensitive to the $\sim gT$ sector, and lead to contact terms  which
correct the collinear emission.

These simplifications are a consequence of the following
physical picture.  Since the hard and jet-particles are propagating 
almost exactly along the light cone, they are probing an essentially
undisturbed plasma, at least as far as the soft classical background
is concerned. 
Thus, the soft correlations that are probed by are
the hard and jet-particles are statistical rather 
than  dynamical, $i.e.$ these are correlations
that arise from the partition function, rather than a retarded
response averaged over the correlated initial conditions determined
by the partition function.   This heuristic explanation 
underlies the  dimensionally-reduced lattice calculation of $\hat q$  
which is described in \sect{chap_lattice}.

At a technical level, these simplifications were first realized
using light cone causality of retarded propagators\cite{CaronHuot:2008ni},
which says that
\st
\label{causality}
G_{R}(p^+, p^-, \p_\perp) = \int dx^+ dx^- d^2 x_\perp \; e^{i (p^+ x^- + p^- x^+ - \p_\perp \cdot \x_\perp) }  G_{R}(x^+, x^-, \x_\perp)
\stp
is an analytic function of $p^+$ in the upper half-plane at fixed $p^-$ 
and $p_{\perp}$. This is  because  the retarded response
function is only non-zero in the forward light
cone $2x^+x^-\ge \xp^2$. Thus the integral in \Eq{causality} has support only for $x^-> 0$,
and the Fourier integral provides an analytic continuation in the upper half $p^+$ plane,
due to the decreasing exponential $e^{ip^+ x^-}$.
To illustrate these simplifications in a concrete context, we will evaluate the drag and 
transverse momentum diffusion coefficients, \Eq{qT} and \Eq{qL},  at leading order.
To do this we will first briefly review Hard Thermal Loops 
in \sect{HTL}. Then we will use the HTL theory to evaluate \Eq{qT} and \Eq{qL} 
in \sect{sub_euclid} and \sect{sub_sumrule}, using  light cone causality to simplify the calculation.

\subsection{Hard Thermal Loops and Light Cone Correlation Functions}
\label{HTL}

To briefly describe the Hard Thermal  Loops consider a scalar field
 $\phi_a$ transforming in the representation $R$
(where $a$ is a color index).  There is a single particle
density matrix $N_{ba}(t, \p, \x)$  transforming in the $R\times\bar R$
representation which
records the statistics of the color
orientations of a quasi-particle excitation. 
Heuristically~\cite{Arnold:1998cy},
$N_{ba}(t,\p,\x)$ is a quasi-local expectation value  of $\llangle a_{a,\p}^\dagger(t,\x) a_{b,\p}(t,\x) \rrangle$ at the space time point $(t,\x)$.
More
formally,  $N_{ba}(t,\p,\x)$ can be defined as the  Wigner
transform of the gauge-covariant Wightman function~\cite{Blaizot:2001nr}.
The gluon singlet distribution function (denoted $n_\p$ above) is proportional to
the color trace of a single particle density matrix,
\st
n_\p \equiv \frac{1}{d_R}  \Tr{N(t,\x,\p) }
\stp
while the color current associated with the gluon distribution is\footnote{The leading factor of $2$ is the spin of the gluon, and for a scalar field
this factor would not appear.}
\st
J^{A,\mu}(t,\x) = 2 g\int \frac{d^3\p}{(2\pi)^3} \Tr{ T^A N(t,\x,\p) } v_\p^{\mu},
\stp
where $v^\mu_\p=(1,\hat\p)$.
The single particle density matrices $N_{ba}$ 
obey a Vlaslov equation
\st
\label{colorboltz}
(D_t + \v \cdot \D_{\x} )  N_\p   
+ \half g \left\{ (\bm E + \v \times \bm B)_i , \frac{\partial N_\p}{\partial \p_i} \right\} = 0 \, ,
\stp
where ${D_\mu}= \partial_{\mu} - i g [A_{\mu}, \cdot ]$ denotes the appropriated gauge covariant derivative
for the $R \times \bar R$ representation, and the curly brackets denote an
anti-commutator.
The particles source the classical chromo-electric and magnetic fields
\st
\label{emequations}
- D_\mu F^{\mu\nu}(t,\x)  = J^{\nu}(t,\x) \, .
\stp

The Hard Thermal Loop retarded Green function is derived from this Maxwell-Boltzmann system of  equations, \Eq{colorboltz}
and  \Eq{emequations}, by expanding around the equilibrium state, $n_\p=1/(e^{E_\p/T} - 1)$
\begin{align}
   N_{ba}  =& n_\p \delta_{ba} + \delta N_{ba} \, , \\
   A_{\mu} =& 0 +  \delta A_{\mu} \, ,
\end{align}
and solving for the response of the classical field to an external
current, 
$A^{\mu}(\omega,\q) = iG_{R}^{\mu \nu}(\omega,\q) J_{\nu,\rm ext}(\omega,\q)$.
We will not go through this set of steps but 
refer to the literature~\cite{Blaizot:2001nr,Litim:2001db}. The resulting retarded Green functions
in the Coulomb gauge are
\begin{subequations}
   \label{htl}
\begin{align}
   G_R^{00} (\omega,\q) =& \left. \frac{-i \eta^{00}}{q^2 + \Pi_{L}(q^0/q)} \right|_{q^0 = \omega + i\epsilon} \, , \\
   G_R^{ij}(\omega,\q) =&  \left. \frac{-i  (\delta^{ij} - \hat q^i \hat q^j)}{- (q^0)^2 + q^2 + \Pi_T(q^0/q) }  \right|_{q^0 = \omega + i\epsilon} \, , 
\end{align}
\end{subequations}
where  
$\eta^{00} =-1$ 
and the self-energies are 
\begin{subequations}
\begin{align}
\Pi_{L}(x) =&  m_{D}^2 \left(1 - \frac{x}{2} \log\left( \frac{x + 1}{x - 1} \right) \right)  \, ,   \\
   \Pi_{T}(x) =&  \frac{m_{D}^2}{2} \left(x^2  - \frac{(x^2 -1) x}{2} \log\left( \frac{x + 1 }{x-1} \right)  \right) \, ,
\end{align}
\end{subequations}
with $x=q^0/q$.
The spectral density is the difference between the retarded and 
advanced green functions
\st
\label{htlrho}
\rho^{\mu\nu}(\omega) = G_{R}^{\mu\nu}(\omega, \q) - G_{A}^{\mu\nu}(\omega, \q) \, , 
\stp
where $G_{A}(\omega,\q)$ is found by setting $q^0 = \omega - i\epsilon$ in
\Eq{htl}. The HTL spectral density and its properties are discussed
in the references~\cite{Blaizot:2001nr}.

As we described above,  $\hat q$ and $\hat q_L$ are determined
by the propagation of the hard and jet particles  
in a random classical background field. In equilibrium, the 
fluctuation-dissipation theorem relates the  symmetrized 
two point function $G_{rr}(\omega,\q)$ to the imaginary part of the retarded response\footnote{Our notation $G_{rr}$ follows the one of the
so-called $r/a$ formalism of Thermal Field Theory. We refer to
this reference \cite{CaronHuot:2007nw} for a comprehensive
set of Feynman rules for the HTL theory in that formalism.}
\st
\label{FDT}
G_{rr}^{\mu\nu}(\omega, \q) \equiv  \frac12 \llangle \{ A^{\mu}(\omega,\q), A^{\nu} \} \rrangle = \left( n_B(\omega) + \frac{1}{2} \right) \rho^{\mu\nu} (\omega,\q) \, .
\stp
For a classical background field $A_{\mu}$, the symmetrization of the field operators is irrelevant, and the symmetrized two point function simply records
the two-point statistics of the classical field. Approximating \Eq{FDT} for
$\omega \ll T$ yields
\st
\label{softprop}
G_{rr}^{\mu\nu}(\omega,\q) = \llangle A^{\mu}(\omega,\q) A^{\nu} \rrangle   =\frac{T}{\omega} \rho^{\mu\nu}_{HTL} (\omega,\q) \, .
\stp
We will use these statistics of the 
classical soft modes when evaluating the longitudinal drag
and diffusion coefficient in the next section.

In addition to these rather complex propagators, the Maxwell-Boltzmann system leads to 
additional HTL vertices which describe how the soft 
classical background interacts with itself. 
The resulting Feynman rules describing this response are rather
complex as well. To date, this elaborate set of Feynman vertices has been 
used to compute  the dilepton emission rate at small invariant mass\cite{Braaten:1990wp}
and the diffusion coefficient of a heavy quark at
NLO\cite{simonguy}.  As 
discussed above and in the next sections, 
these intricate HTL rules simplify dramatically when evaluating the lightcone correlators involved in energy loss. 
Indeed, the NLO computations of energy 
loss described in \sect{sec_nlo} would not have been possible without these
simplifications.

\subsection{Transverse diffusion and Euclidean operators}
\label{sub_euclid}
In this section we will compute  $\hat q$ and $\cc(\qp)$ at LO using
Eqs.~\eqref{qT}, \eqref{defcq} and \eqref{cfromw} as our starting point. Recall
that these Wilson line 
correlators describe the propagation of a hard or jet particle
traversing the classical field provided by the soft modes. The
statistics of these classical fluctuations are given by Eqs.~\eqref{FDT}
and \eqref{softprop}.
As we will see, light cone causality \Eq{causality} dramatically simplifies the calculation, leading to a Euclidean formulation.

\begin{figure}[ht]
\begin{center}
	\includegraphics[width=8cm]{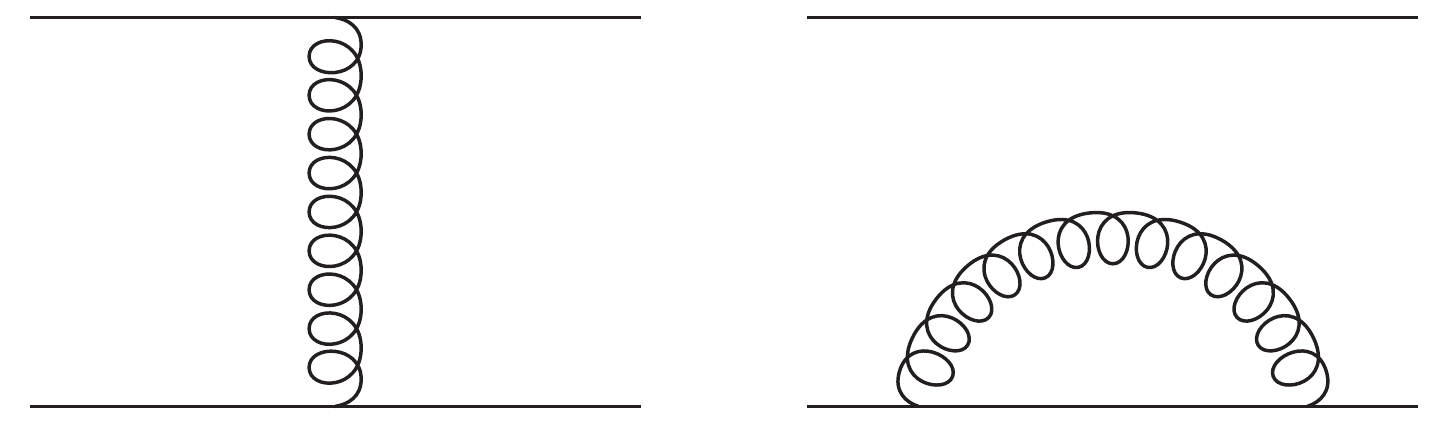}
\end{center}
\caption{Diagrams for the leading-order evaluation
of $\cc(\xp)$ in a non-singular gauge. The plain lines
represent the Wilson lines along the light-cone, at transverse separation
$\xp$. The diagram with the self-energy on the upper Wilson line is not shown.}
\label{fig_lo_cq}
\end{figure}

For $\cc$, applying the definition~\eqref{cfromw}, one finds (in any non-singular gauge)
that the diagrams
in Fig.~\ref{fig_lo_cq} contribute.
Clearly, the diagram on the right and its symmetric give rise to
the $\xp$-independent part in Eq.~\eqref{defcb}, whereas the one on the left
gives rise to the $\xp$-dependent exponential, so that,
at leading order, for an adjoint source (a gluon) one has
\begin{eqnarray}
\label{C_LO_unevaluated}
\cc(\qp) &=&g^2\ca\int\frac{dq^0dq_z}{(2\pi)^2}2\pi\delta(q^0-q_z) G_{rr}^{--}(Q),\\
\label{C_LO_xp_unevaluated}
\qquad \cc(\xp)&=&\int\frac{d^2\qp}{(2\pi)^2}\left(1-e^{i\bxp\cdot\bqp}\right)\cc(\qp),
\end{eqnarray}
where $G_{rr}^{--}(Q)=v_\mu v_\nu G^{\mu\nu}_{rr}(Q)$ is given
for soft momenta  by Eq.~\eqref{softprop}. Since
$q^0=q^z$ enforces space-like kinematics, it would then
seem that one is left with an integral of the non-trivial
HTL spectral function \Eq{htlrho} over the entire Landau cut, $\vert q^0\vert<q$. 
Such integral poses no particular
challenge numerically,\cite{Arnold:2001ms} but any extension beyond leading order, with loops composed
of HTL propagators and vertices generated by the rules discussed in the previous
section~\ref{HTL}, would be extremely challenging from the computational
viewpoint. We refer to the computation to NLO\cite{simonguy}  of $\kappa$, the analogue of $\qhat$ for
a very massive quark, for an example of such an intricate calculation. 
In the present case, however, there is a considerable simplification. As we
have remarked in \sect{chap_classical}, $\cc(\qp)$ can be 
obtained from a Wilson loop supported on a light front (a null plane),
as show in Fig.~\ref{wilsonfig}.
The special analiticity properties of objects of this kind are the
key to their simplification.
Indeed, Alves, Das and Perez,\cite{Alves:2002tx} and Weldon,\cite{Weldon:2003uz,Weldon:2003vh}
introduced the technique of light-front
quantization into Euclidean Thermal Field Theory, extending the usual imaginary time
formalism from the space-like domain to light fronts.
Subsequently, and independently, Caron-Huot \cite{CaronHuot:2008ni} introduced a similar quantization
technique and showed how it can reduce the calculation of (the soft contribution
to) this Wilson loop to a much simpler Euclidean calculation
in dimensionally-reduced EQCD. 

The first step in this endeavor consists in showing that 
the calculation of $n$-point correlation functions, where all
fields lie on a spacelike hypersurface, can be carried out by Euclidean
techniques \cite{Alves:2002tx,Weldon:2003uz,CaronHuot:2008ni}. 
The null correlators we need can
also be computed provided that they are free of collinear singularities,
which they are up to NLO included.\cite{CaronHuot:2008ni} The proofs presented in
Refs.~\cite{Alves:2002tx,Weldon:2003uz,CaronHuot:2008ni} are
rather complex and technical. 
Here we will present a much simpler
derivation, also due to Caron-Huot%
\footnote{S.\ Caron-Huot, oral presentation at the Institute for Nuclear
  Theory (Seattle), 29 March 2012},  already 
collected in~\cite{Ghiglieri:2013gia}~, which we follow closely.
It is an illustration of the two-point function case,
which, as one can see from Eq.~\eqref{C_LO_unevaluated}, will nicely yield the LO result for 
$\cc(\qp)$.

Consider the symmetrized correlator of some operator (such as $G^{--}$ 
in our case), $G_{rr}(x^0,\x)$ with $\vert x^z\vert  > \vert
x^0\vert$.  Since the separation is spacelike, operators commute, and
therefore $G_{rr}$ equals $G^<$, or $G^>$, as shown in
\Eq{softprop}.  We are interested in the lightlike limit $x^z= x^0$.
There are additional subtleties in taking this limit, but these
are irrelevant for the soft and leading order contributions to
such correlators. 
Let us write $G_{rr}$ in terms of its Fourier representation, i.e. the inverse
of \Eq{causality}
\begin{equation}
	G_{rr}(x^0,\x) = (2\pi)^{-4} \int d\omega \int dp^z d^2 p_\perp
e^{i(x^z p^z+\bxp\cdot \bpp - \omega x^0)}\:
G_{rr}(\omega,p^z,p_\perp),
\end{equation}
and use the exact relation between the spectral function and 
the symmetric propagator, \Eq{FDT}. 
We then shift one integration variable to $\tilde p^z = p^z - (x^0/x^z) \omega$:
\begin{equation}
\label{almostthere}
G_{rr}(x^0,\x) = \int \frac{d\omega d\tilde{p}^z d^2 p_\perp}{(2\pi)^4}
e^{i( x^z \tilde p^z + \bxp \cdot \bpp)}
\left(\nbe(\omega)+\frac12\right)\: 
\rho(\omega,\tilde{p}^z+\omega(x^0/x^z),p_\perp) \,,
\end{equation}
so that the Fourier exponent is now frequency-independent and we can
perform the $\omega$ integration by contour methods. 
The spectral weight is written as
\st
\rho(\omega, \p)
=  G_{R}(\omega, \p) - G_{A}(\omega, \p) \, ,
\stp
and  the integral over $G_{R}(\omega,p)$ is performed
by closing the contour in the uppper half plane,  while the integration
of $G_{A}(\omega,p)$ is performed by closing the contour in the lower half plane.
Since $|x^0/x^z|<1$,  
$G_R(\omega,\tilde{p}^z+\omega(x^0/x^z),p_\perp)$ 
is an analytic function of $\omega$ in the upper complex $\omega$ plane.
This is a consequence of  (slightly generalized) light-cone causality%
\footnote{ 
   This
   is precisely a statement of lightcone causality as described in \Eq{causality} when $x^0/x^z=1$.
   In this case,
$\tilde p^z = -p^-$ and $\omega$ parametrizes the
 analytic continuation in $p^+$  at fixed $p^-$ and $p_\perp$. }
as described in and after \Eq{causality}.
The advanced
function is similarly an analytic function in the lower complex $\omega$ plane.
Therefore the only singularities encountered in continuing the frequency
integration are those in the statistical function $(\nbe(\omega)+1/2)$,
which has poles at $\omega_n = 2\pi inT$ with $n=(\ldots
-1,0,1,\ldots)$ and residues equal to $T$. 
Closing the contour around 
these poles, and renaming
$\tilde{p}^z$ to $p^z$, we find%
\footnote{
  The pole at $n=0$ is an artifact of the separation of $\rho$, which vanishes for $\omega=0$, into $G_R$ and $G_A$.  
The individual poles there can then be dealt with in a principal value prescription, for instance.
}
\begin{equation}
\label{simonmagic}
G_{rr}(x^0,\bx) = T \sum_{n} \int \frac{d^3 p}{(2\pi)^3}\,
e^{i \x \cdot \p}\,
G_E\left(\omega_n,p^z+i\omega_n \frac{x^0}{x^z},p_\perp\right) , \quad
\omega_n = 2\pi nT \,,
\end{equation}
where we have recalled 
that the retarded 
green function is determined by an analytic continuation
of the Euclidean function, $G_R(\omega,p) = -i G_E(\omega_E=-i(\omega+i\epsilon),p)$.

When  we need to compute the soft $gT$ contribution to such
a correlator, one may drop the nonzero Matsubara frequency
contributions; that is, we keep only the $n=0$ term in the sum.  
This term,
$G_E(\omega_n,p^z+i\omega_n (x^0/x^z),p_\perp) \to G_E(0,p^z,p_\perp)$,
is the Euclidean correlation function of the 3-dimensional dimensionally
reduced  EQCD theory, which we shall describe in more detail
in Sec.~\ref{chap_lattice}. Thus, the soft contribution to Euclidean operators
at spacelike and lightlike separations is time-independent.

Let us go back to the computation of $\cc(\xp)$. As all points in that Wilson loop
are either at a space-like or light-like separation, the above introduced
EQCD reduction is applicable. Hence, at leading order, 
Eq.~\eqref{C_LO_unevaluated} becomes
\begin{equation}
\label{C_LO}
\cc(\qp) =g^2\ca T\int\frac{dq_z}{(2\pi)}2\pi\delta(q_z) G_{E}^{--}(0,q_z,\qp)\
=g^2\ca T\left(\frac{1}{\qp^2}-\frac{1}{\qp^2+\md^2}\right),
\end{equation}
The result
is easily understood as the difference between the massless, transverse 
and massive, longitudinal EQCD propagators, which also arise as the $q^0=q^z=0$
limit of \Eq{htl}. For (the soft contribution to) 
$\qhat$ we then have the well-known logarithmic dependence on the cut-off, which
we anticipated in \Eq{eq:qlosummary} in Sec.~\ref{chap_classical}, i.e.
\begin{equation}
\label{q_LO}
\qhat(\mu)=g^2\ca T\int^\mu\frac{d^2\qp}{(2\pi)^2} \frac{\md^2}{\qp^2(\qp^2{+}\md^2)}
=\frac{g^2\ca T\md^2}{2\pi}\ln\frac{\mu}{\md}\,.
\end{equation}

Furthermore, we remark that Eq.~\eqref{C_LO} was already obtained\cite{Aurenche:2002pd}
using a sum rule. Caron-Huot has shown\cite{CaronHuot:2008ni}
that the two approaches are equivalent, and that the sum rule is possible
because of the aforementioned analyticity of retarded and advanced functions. 
These very same properties will be of great importance for the
evaluation of non-Euclidean operators.

Finally, let us consider the effect of the non-zero Matsubara modes
we have neglected in Eq.~\eqref{C_LO}. If $\qp \sim gT$ is kept soft,
the subtleties
in putting $x^0/x^z=1$ will 
need to be reexamined when 
including
the non-zero Matsubara modes in the sum. 
The evaluation of 
these contributions  is related
to the collinear singularities that should show up at NNLO. It would
be interesting to clarify these collinear singularities with direct
calculations.
If one considers shorter transverse separations (with $\qp\sim T$)
then all Matsubara modes will contribute straightforwardly, and 
the Euclidean light-front computation in Eq.~\eqref{simonmagic} may 
 be more cumbersome than a direct computation in the real-time formalism.
The leading-order contribution to $\qhat$ from the scale
$T$ has been computed\cite{Arnold:2008vd} using real time methods.

\subsection{Longitudinal diffusion and non-Euclidean operators}
\label{sub_sumrule}
\begin{figure}[ht]
\begin{center}
\includegraphics[width=5cm]{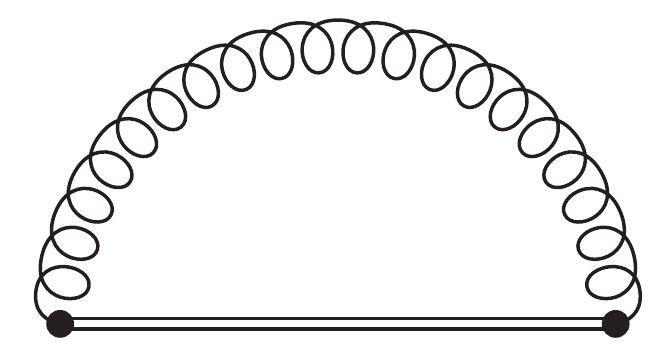}
\end{center}
\caption{The leading-order soft contribution to $\ql$. The two dots
	are the two field strengths and the double line is the adjoint
	Wilson line connecting them.
The curly line is a soft HTL gluon.}
\label{fig_lo_soft}
\end{figure}

As we mentioned at the beginning of \sect{chap_sumrule},
not all lightcone or light-front supported 
operators admit a three-dimensional, Euclidean description for the soft
modes. A prime example is the longitudinal momentum diffusion coefficent
$\ql$, as given by Eq.~\eqref{qL}. At leading order it is given by
the diagram shown in Fig.~\ref{fig_lo_soft}.
In any non-singular gauge  it reads
\begin{equation}
\label{lo}	\ql=g^2\ca \int_{-\infty}^{+\infty}dx^+
\int\frac{d^4Q}{(2\pi)^4}e^{-iq^-x^+}(q^+)^2G^{--}_{rr}(Q),
\end{equation}
where  again $G^{--}_{rr}(Q)$ is given by Eq.~\eqref{softprop}.  
The $x^+$ integration sets $\qm$ to zero. 
We clearly see that, although
originating from a lightcone operator, $\qll$ cannot be evaluated in EQCD:
indeed, the zero-mode contribution exactly vanishes when the previous
techniques are applied.

\begin{figure}
\begin{center}
\includegraphics[width=9cm]{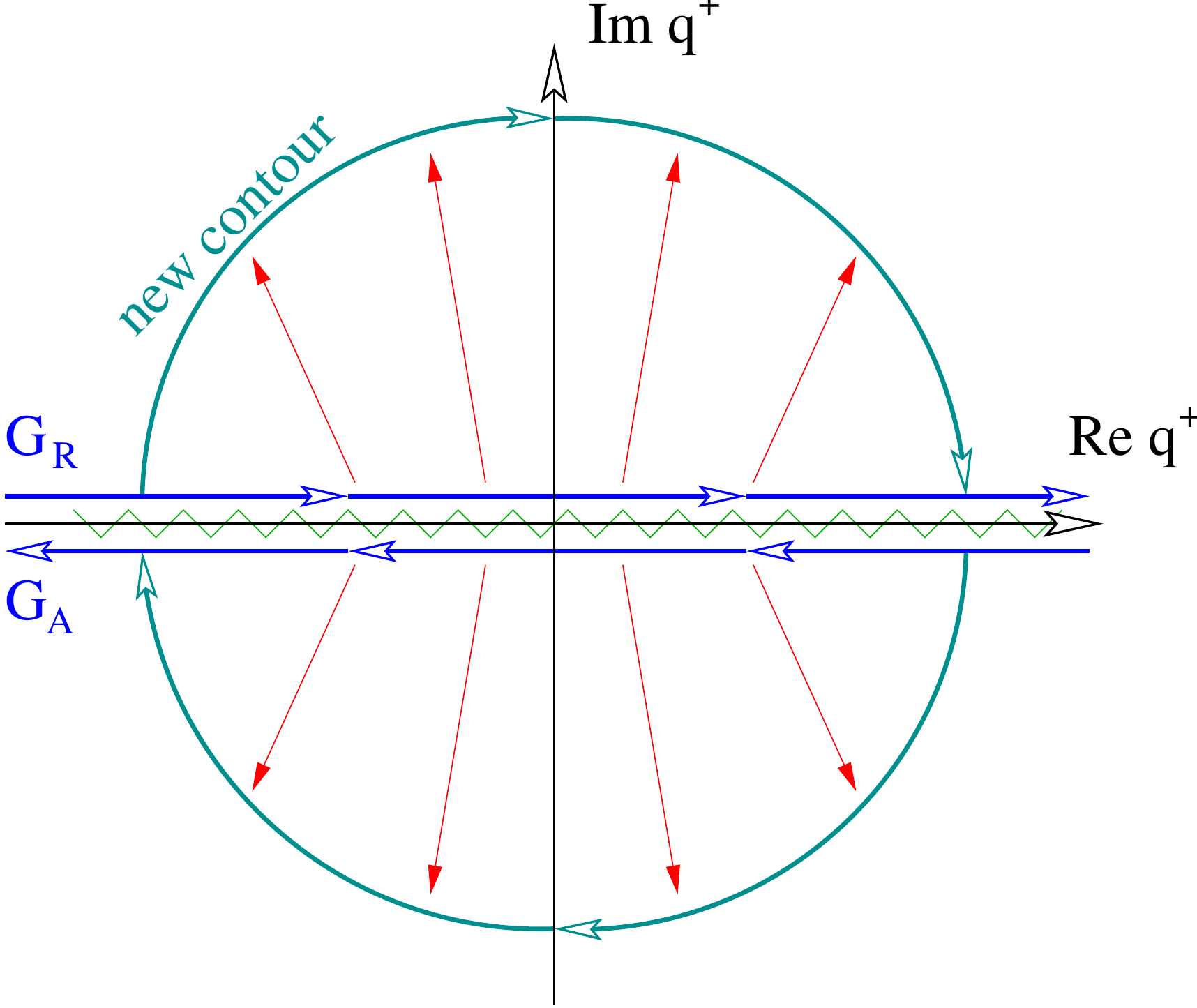}
\end{center}
\caption{\label{fig_contour}  Integration contour in the complex $q^+$
integration, and the deformation we use to render $q^+ \gg gT$. $G_R$ 
runs above the real axis and $G_A$ below. }
\end{figure}

We can however evaluate Eq.~\eqref{lo} by employing sum rules
that are rooted in the same analyticity properties that were used in
the derivation of Eq.~\eqref{simonmagic}. In detail,
we plug the result of
Eq.~\eqref{softprop} in \Eq{lo}. 
Up to $\OO(g^2)$ corrections\footnote{When expanding
	the statistical factor in the soft region in Eq.~\eqref{softprop},
one has $\nbe(\omega)+1/2=T/\omega(1+\OO(g^2))$.}
we then have
\begin{equation}
\label{lo2}	\ql=g^2\ca
\int\frac{dq^+d^2\qp}{(2\pi)^3}Tq^+ (G^{--}_R(q^+,\qp)-G^{--}_A(q^+,\qp)).
\end{equation}
This too would be a simple enough numerical integral\cite{Braaten:1991jj} over the HTL
spectral function in the Landau cut, of difficult extension to higher orders.
However, as we have previously remarked, retarded (advanced)
two-point functions are analytic in the upper (lower) half-plane
in any timelike or light-like variable. We can thus 
deform the integration contours\cite{jetwip} away from the real axis onto $\calr$
($\vert\qll\vert=\mu^+\gg gT$, $\mathrm{Im}\,\qll>0$)
and $\cala$ ($\vert\qll\vert=\mu^+\gg gT$, $\mathrm{Im}\,\qll<0$), 
as depicted in Fig.~\ref{fig_contour}.\footnote{The longitudinal and transverse
contributions to $G_R^{--}(Q)$ contain poles at $\qll=\qm/2\pm i\qp$ ($q^2=0$),
which, being on both sides of the complex plane, appear to 
violate analyticity. However their residue cancels in the sum of 
longitudinal and transverse components. As observed in \cite{CaronHuot:2008ni}~,
they are artifacts of the decomposition into Lorentz-variant
longitudinal and transverse modes
and their contribution has to vanish in all gauge-invariant quantities.}
$\mu^+$ is a UV cutoff on $q^+$, to make sure that we consider only the soft 
region. As we shall see, it will have no effect.

Along the arcs $\qll$ is then large, albeit complex.  
The longitudinal and transverse propagators simplify greatly, i.e.
\begin{eqnarray}
G^{--}_R(q^+,\qp)&=&G^L_R(q^+,\qp)+\frac{\qp^2}{q^2}
 G^T_R(q^+,\qp)\\
\label{arcexpand}
 &=&
\frac{i}{(q^+)^2}+\frac{\qp^2}{(\qll)^2}
\frac{-i}{\qp^2+\mmg}+\order{\frac{1}{(\qll)^3}},
\end{eqnarray}
where we have used the Coulomb gauge propagators given in \Eq{htl} and
$\mmg=\md^2/2$ is the LO gluon asymptotic thermal mass as given
in Eq.~\eqref{eq:asymptoticm}.  What this equation
is telling us is that, for large (complex) $\qll$, the longitudinal part
is unchanged with respect to its unresummed version, whereas the transverse
part acquires the asymptotic mass, which is typical of collinear, transverse
excitations. 

Upon plugging Eq.~\eqref{arcexpand} in Eq.~\eqref{lo2}, the
$\qll$ integration along the arcs is trivial and the end result is\cite{jetwip} 
\begin{equation}
\label{lofinal}	\ql(\mu)=g^2\ca T
\int^\mu\frac{d^2\qp}{(2\pi)^2}\frac{\mmg}{\qp^2+\mmg}
=\frac{g^2\ca T }{2\pi}\mmg\ln\frac{\mu}{m_\infty},
\end{equation}
where contributions smaller than $1/(\qll)^2$ in Eq.~\eqref{arcexpand} are not needed,
as they would only give rise to power-law terms in the cutoff on 
$\qll$, which would then cancel
against contributions from larger scales.
The transverse integration has been regulated in the UV with $\mu\gg gT$,
as in Eq.~\eqref{q_LO}. Again, this result had been
anticipated in \Eq{eq:qlosummary}. As we commented after \Eq{eq:md},
this calculation, together with the one in the previous subsection, clearly
shows the physical reasons causing the Debye mass $\md$ to appear in $\qhat(\mu)$
and the asymptotic mass $m_\infty$ to appear in $\ql(\mu)$.

The sum rule we have just illustrated has been derived in this fashion
in\cite{jetwip}~. It can also be seen as the bosonic equivalent of the one presented
in the calculation of the thermal photon rate to NLO\cite{Ghiglieri:2013gia}
and of the  thermal ultrarelativistic right-handed neutrino rate at LO\cite{Besak:2012qm}. 
We also remark that the same result~\eqref{lofinal}
was previously obtained in a different way\cite{Peigne:2007sd}
for the drag coefficient $\hat{e}$, which is related by the Einstein 
relation \Eq{einsteinrel}.
As it is shown there\cite{Peigne:2007sd}, once the difference in regularization
between  $\qp<\mu_{\perp}$ and $q<\mu_q$ is taken into account, 
Eq.~\eqref{lofinal} agrees with the
numerical results of Braaten and Thoma\cite{Braaten:1991jj} for $v\to1$.

In summary, in this Section we have introduced Hard Thermal Loops
and their simplification on the light cone. In \sect{sub_euclid}
we have analyzed the subset of operators, including transverse 
momentum diffusion, that admit a three-dimensional, Euclidean
description in the soft region, whereas in the current subsection
we have analyzed those that do not, showing how they still considerably
simplify. In the following Section we will show how Euclidean
operators can be calculated on the lattice, whereas in \sect{sec_nlo}
we will use the results of this Section to generalize the perturbative
kinetic approach to NLO.

\section{Non-perturbative determinations of $\cc(\xp)$ and $\qhat$}
\label{chap_lattice}

As we mentioned previously, the field-theoretical developments
introduced by Caron-Huot \cite{CaronHuot:2008ni} make it possible
to map certain operators supported on light fronts and light cones
to the Euclidean domain, where the soft contribution is dominated
by the dimensionally reduced zero modes of 3D EQCD.
Recently, this has been exploited to compute
the soft contribution to $\cc(\xp)$ and $\qhat$ on the lattice.
Laine and Rothkopf \cite{Laine:2013lia} have performed a first
study of the Wilson loop defined in Fig.~\ref{wilsonfig} 
in classical gauge theory,
which allows  real-time lattice studies and catches the qualitative nature
of the soft physics,
which, has we have argued in Sec.~\ref{chap_classical}, is of classical origin. In particular, they
also examined the dependence on $v$ of the Wilson loop (i.e. whether the 
approach to the light-cone from above or below is different for classical fields), 
finding it modest.

Panero, Rummukainen and Sch\"afer \cite{Panero:2013pla} have instead
computed the soft contribution to $\cc(\xp)$ and to $\qhat$ by means of
lattice simulations of EQCD.\footnote{
	A different method, not based on the properties 
	of Sec.~\ref{chap_sumrule}, for estimating $\qhat$ on the lattice
	has been proposed\cite{Majumder:2012sh}.}
 In the remainder of this Section, we will
illustrate the basic principles of this treatment and its potential application
to other light-cone observables that are suited to it, such as 
$\qhat(\delta E)$, which will be defined in Sec.~\ref{sec_nlo}, or the asymptotic masses. 
To this end, we start by reviewing Electrostatic QCD (EQCD) 
\cite{Braaten:1994na,Braaten:1995cm,Braaten:1995jr,Kajantie:1995dw,Kajantie:1997tt},
which is a dimensionally-reduced EFT of QCD describing the
time-independent, long-wavelength $\lambda\sim 1/(gT)$ modes of the latter theory.
If treated perturbatively, it breaks down at very long wavelengths (very
small momenta) $\lambda\sim 1/(g^2T)$ ($k\sim g^2T$) where 
magnetic non-perturbative contributions\cite{Linde:1980ts,Braaten:1994na}
appear. If, on the other hand, it is treated 
non-perturbatively, both the electric and magnetic scales are treated correctly.
Hence, a lattice calculation in EQCD, such as the one of $\qhat$ we are considering, 
determines the soft ($gT$)
and ultrasoft ($g^2T$) contributions. Indeed, nonperturbative EQCD 
has been employed for many studies of the thermodynamics of QCD, as in this example\cite{Hietanen:2008tv},
and of the gauge
sector of the standard model, as in this review\cite{Laine:2000xu} and this recent 
work\cite{Laine:2012jy}.

The Lagrangian of EQCD reads	
\begin{equation}
\label{EQCDlag}
\mathcal{L} = \frac{1}{4} F_{ij}^a F_{ij}^a + \tr \left( (D_i A_0)^2 \right) + \mE^2 \tr \left( A_0^2 \right) + \lambda_3 \left( \tr \left( A_0^2 \right) \right)^2 ,
\end{equation}
where only operators of dimension up to three have been kept. 
$D_i=\partial_i-i g_E[A_i,\cdot]$ is the 
standard covariant derivative, so that the Lagrangian describes a 3D SU(3)
Yang-Mills field minimally coupled to an adjoint scalar field $A_0$, which
has a mass and a quartic coupling (adjoint Higgs). Fermions,
having only odd Matsubara frequencies proportional to $\pi T$, are completely
integrated out in the dimensional reduction.

At leading order the matching between QCD and EQCD yields 
\cite{Braaten:1994na,Braaten:1995jr,Kajantie:1997tt}
\begin{equation}
	g_E^2=g^2(\mu)T,\qquad \mE^2=\left(\frac{\nc}{3}+\frac{\nf}{6}\right)
	g^2(\mu)T^2,\qquad \lambda_3=\frac{9-\nf}{24\pi^2}g^4(\mu)T,
	\label{lomatching}
\end{equation}
where $g_E$ is the dimensionful coupling of EQCD and $g(\mu)$ is the
scale-dependent coupling of QCD. The mass parameter is equal to the
leading-order Debye mass. The positive dimensionality of $g_E$ makes the theory
super-renormalizeable.

The EQCD action can be readily discretized for lattice studies 
We refer to \cite{Hietanen:2008tv,Panero:2013pla} for details on this procedure.
The implementation of the Wilson loop 
\cite{CaronHuot:2008ni,D'Eramo:2010ak,Benzke:2012sz} depicted in Fig.~\ref{wilsonfig}
is on the other hand somewhat non-trivial  in the dimensionally-reduced theory, where
the time dimension is integrated out. In the continuum, the 
Wilson line along the light-cone direction changes from 
\begin{equation}
	\label{wlineqcd}
	U(x^+_f;x^+_i)=\mathcal{P}\exp\left[-ig \int_{x^+_i}^{x^+_f}
	dx^+\big(A^0(x^+)-A^z(x^+)\big)\right]\;\text{in Minkowskian QCD,}
\end{equation}
to
\begin{equation}
U(x_{z,f};x_{z,i})=\mathcal{P}\exp\left[-ig \int_{x_{z,i}}^{x_{z,f}}
dx_z\left(iA_0(x_z)-A_z(x_z)\right)\right]\;\text{in EQCD},
	\label{wlineeqcd}
\end{equation}
where the $A^0$ in Eq.~\eqref{wlineqcd} is the temporal gauge field of Minkowskian
4D QCD, whereas the one in Eq.~\eqref{wlineeqcd} is the adjoint scalar of Euclidean, 
3D EQCD, the extra $i$ coming from the Wick rotation. The other coordinates
($\bxp$ and, in Eq.~\eqref{wlineqcd}, $x^-$) are understood to be constant and are not shown.

\begin{figure}[ht]
	\begin{center}
		\includegraphics[width=10cm]{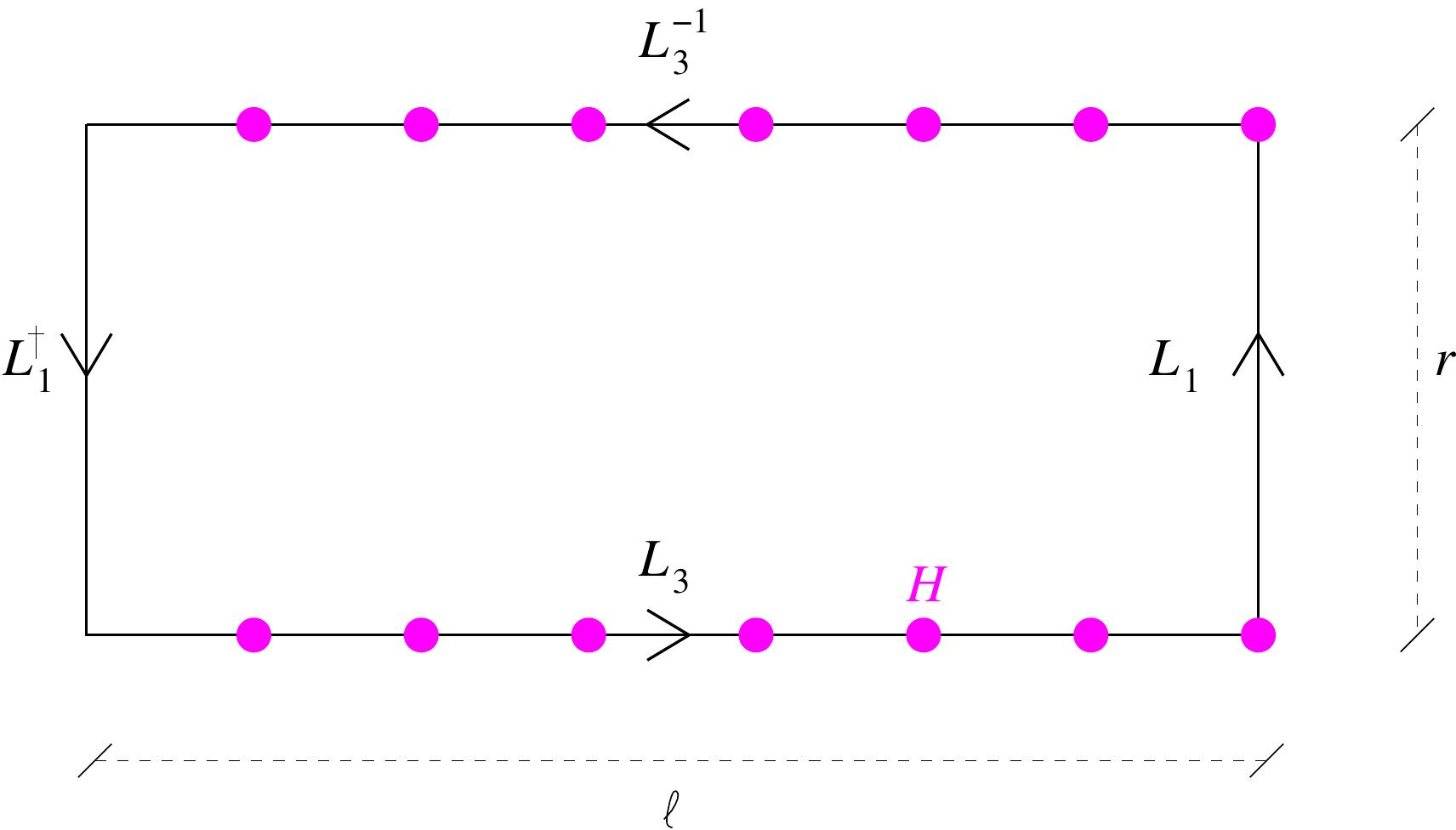}
	\end{center}
	\caption{The Wilson loop yielding $\cc(\xp=r)$ in
		EQCD. The dots on the lines in the horizontal
		direction represent the insertion of the Hermitian operator
		$H$ at the lattice sites, whereas the lines connecting
		them are the gauge links in the $z$ direction.
		Figure taken from the original reference \cite{Panero:2013pla}.}
	\label{fig:lattice_wloop}
\end{figure}

The lattice version of Eq.~\eqref{wlineeqcd} is also non-trivial, because
in the discretized action the 3D gauge fields are naturally re-expressed as gauge
links between the lattice sites, while the $A_0$ scalar field is defined
on the lattice sites. This leads to, in the notation of \cite{Panero:2013pla}
\begin{equation}
L_3( x, a n_\ell ) = \prod_{n=0}^{n_\ell -1} U_3\left( x + a n \hat{3}\right) H \left( x + a (n+1) \hat{3} \right),
\label{latticewline}
\end{equation}
where $L_3$ is the lattice equivalent of Eq.~\eqref{wlineeqcd}, with $n_\ell = \ell/a$
the length of the Wilson line in units of the lattice spacing $a$. $U_3$ is a gauge
link in the $z$ direction and $H(x)$ is a Hermitian, rather than 
unitary, matrix obtained by exponentiation of $A_0(x)$, i.e. 
\begin{equation}
\label{H_definition}
H(x)=\exp[- a g_E^2 A_0(x)] .
\end{equation}
Note that $H(x)$ represents a parallel transporter along a \emph{real-time} interval of length $a$. The transverse Wilson lines $L_1$ are instead the usual product
of gauge links, so that, schematically, the Wilson loop takes the 
form depicted in Fig.~\ref{fig:lattice_wloop}. 

\begin{figure}
	\begin{center}
		\includegraphics[width=6.2cm]{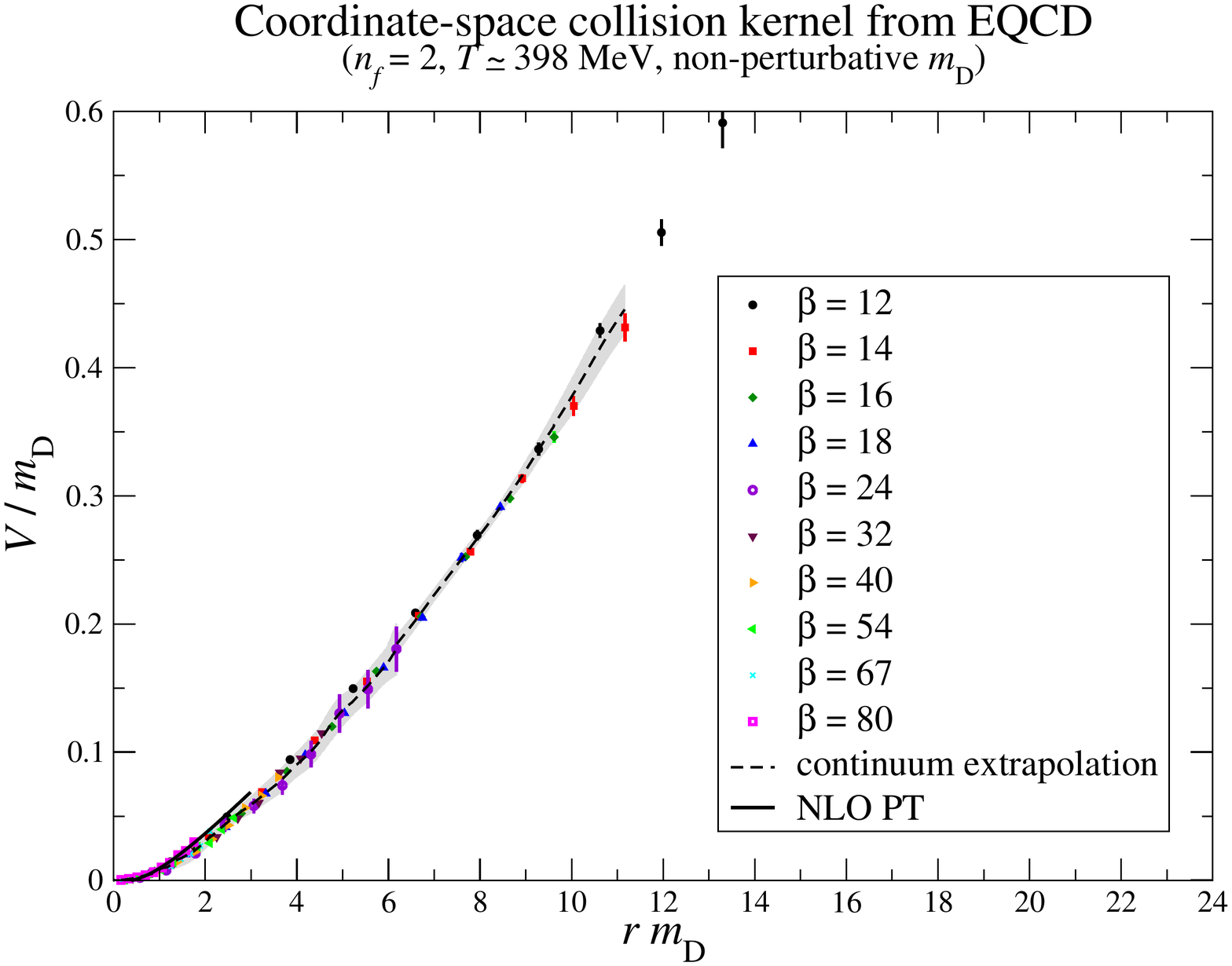}
		\includegraphics[width=6.2cm]{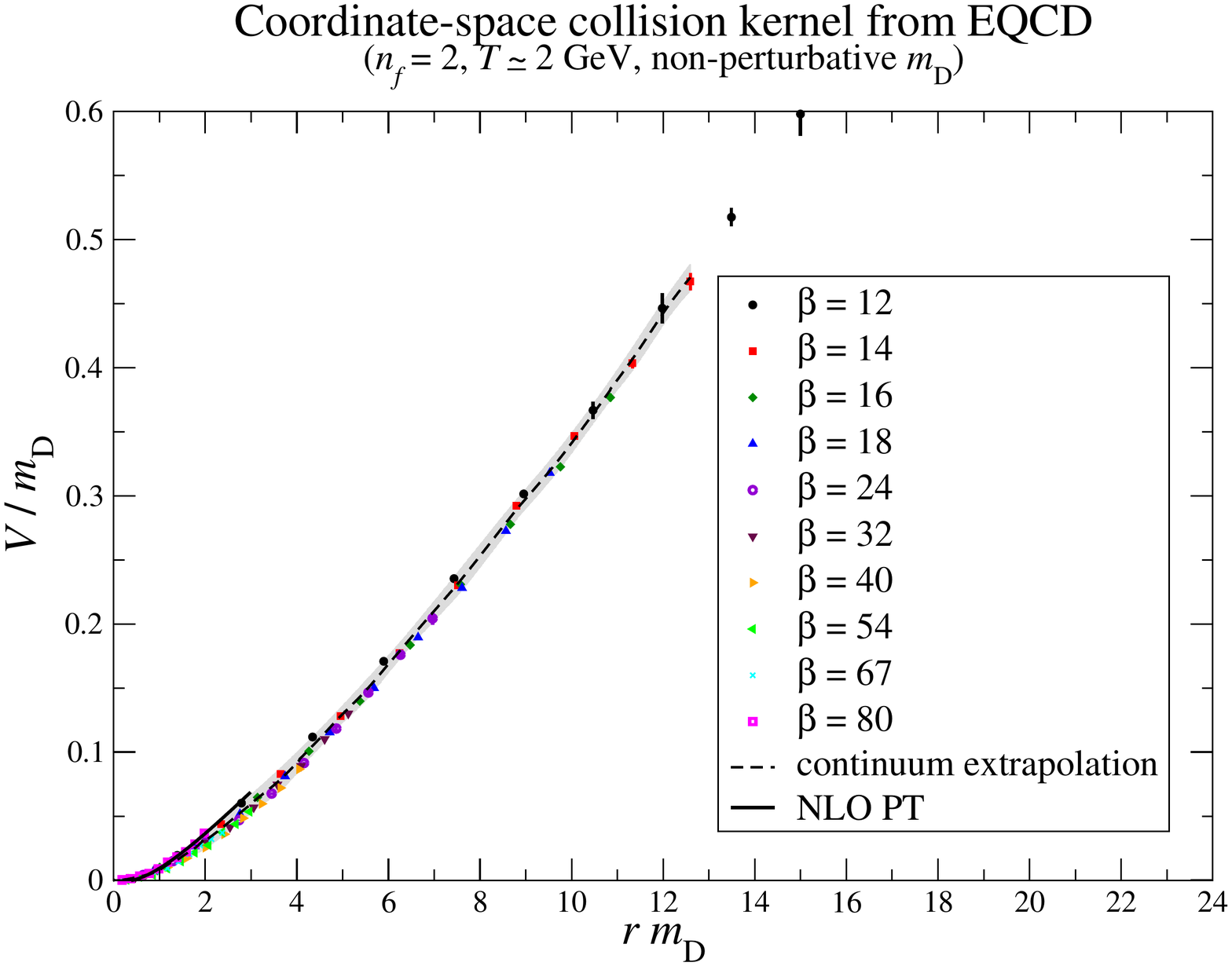}
	\end{center}
	\caption{Results for $\cc(\xp)$ at two different values of the 
		temperature. The Debye mass used as a scale for the axes 
		and plugged in the NLO perturbative calculation 
		\cite{CaronHuot:2008ni} is the non-perturbative one 
		\cite{Laine:1999hh}. Figure taken from the original
		reference\cite{Panero:2013pla}.}
	\label{fig_marco}
\end{figure}

The lattice measurements\cite{Panero:2013pla} of $\cc(\xp)$ through
the above definition are summarized in Fig.~\ref{fig_marco}. As the
authors remarked, once the value the Debye mass which includes
the non-perturbative relative $\OO(g)$ correction\cite{Laine:1999hh}
is used to set the scale and as input for
the perturbative calculation of Caron-Huot, the agreement between
lattice and perturbation theory becomes surprisingly good. If, on the other 
hand, the perturbative Debye mass is used, perturbation theory yields 
a significantly smaller $\cc(\xp)$ at intermediate distances.

The soft contribution to $\qhat$ can then be extracted from the short-distance
behavior of $\cc(\xp)$ along the line of Eq.~\eqref{shortwloop}, i.e.
\cite{CaronHuot:2008ni,Laine:2012ht,Ghiglieri:2013gia}
\begin{equation}
	\cc(\xp)\stackrel{\xp\md\ll 1}{\to} c_1 \xp +\qhat_{\mathrm{soft}}
	\frac{\xp^2}{4}+\OO(\xp^3).
	\label{shortdist}
\end{equation}
The $\qhat_\mathrm{soft}$ extracted in this way also contains a residual
logarithmical dependence on $\xp$, reflecting the leading-order UV
divergence of $\qhat_\mathrm{soft}$ (see Eq.~\eqref{q_LO})
that is absorbed by the hard contribution.
We refer to \cite{Laine:2012ht,Panero:2013pla} for further
details on the matching procedure and the addition of the
leading-order hard contribution \cite{Arnold:2008vd}. Once this has
been performed, Panero \emph{et al.} report an estimate for
$\qhat$ of 6 $\mathrm{GeV}^2/\mathrm{fm}$ at their lower temperature
of 398 MeV, with an estimated uncertainty of 15 to 20\%. We
remark that higher-order contributions from the hard scale,
as well as the possible collinear contribution mentioned
in Sec.~\ref{sub_sumrule}, are absent from this determination.

A possible limitation to the approach we have outlined is that
it relies on a separation of scale between the hard particles,
with momenta of order $\pi T$, and the soft and ultrasoft fields,
characterized by $gT$ and $g^2T$. However, as the authors remark,
the literature suggests
(see for instance \cite{Kajantie:2002wa,Hietanen:2008tv,Laine:2005ai} ) 
 that analytical computations relying
on this separation of scales may be sufficiently accurate 
down to low temperatures, perhaps, suprisingly, down 
to\cite{Borsanyi:2012ve} $T\sim 2 T_c$.

It is worth remarking that the NLO perturbative calculation
predicts at the origin a negative linear slope,\footnote{Due to the
super-renormalizability of EQCD, each loop order causes a different power-law behavior
for $\cc$, so that higher orders cannot contaminate this effect.} i.e. $c_i<0$ in
Eq.~\eqref{shortdist}, which is not
observed in the lattice calculation. This can be attributed to 
discretization errors, which are more severe at short distances,
corresponding to the UV region $\pp\gg \md$ in momentum space. At 
leading order in PT, the dominant UV behavior ($1/\pp^2$) 
cancels between the 
longitudinal and transverse one-gluon exchanges, as shown 
in Eq.~\eqref{C_LO}, leaving a $\md^2/\pp^4$ correction. This
cancellation, while exact in the continuum, is only approximate
on the lattice. D'Onofrio, Kurkela and Moore \cite{D'Onofrio:2014qxa}
have estimated the
associated error to be of order $a/\xp$ and hence especially relevant 
at short distances. Their computation of the renormalization
properties of Eq.~\eqref{latticewline} to order $a$ can help
alleviating these discretization effects.

We also note that, by comparing their results
with the estimate of the magnetic contribution only to $\qhat$
\cite{Laine:2012ht,Benzke:2012sz}, which can be performed similarly
by simulating a Wilson loop in the pure 3D gauge theory, Panero
\emph{et al.} have been able to establish that the magnetic contribution
($g^2T$ only) is subleading compared to the full non-perturbative 
EQCD determination, which includes the $gT$ and $g^2T$ contributions.

Future directions for this very promising new direction can include more
refined measurements of $\qhat$, for instance improving the approach
to the continuum at short distances along the lines discussed before or 
extending the range of explored temperatures. An altogether similar approach
can be used to determine non-perturbatively the soft contribution to other
operators that are amenable to a Euclidean treatment, such as $\qhat (\delta E)$
or the condensates appearing in the definition of the asymptotic masses (see Sec.~\ref{sec_nlo_coll}). One could then envisage for the future the establishment of a factorization
program for the computation of jet-related observables: 
the soft physics, where pQCD struggles, is encoded into these effective operators
measured on the lattice, whereas the harder scales are treated perturbatively.\footnote{
A factorized approach sharing the same spirit, but using holographic methods 
for the description of the medium, has recently been presented in 
\cite{Casalderrey-Solana:2014bpa}~.} EFT
techniques could be used to make the factorization well defined and deal with possible
large logarithms.  A first step in this direction has been taken in this study 
\cite{Brandt:2014uda}. It compares the screening masses
of a specific correlator, extracted 
from lattice measurements, with their determination through a differential equation
that is very similar to Eq.~\eqref{bspace}, up to a complex phase for $\cc(\xp)$.
By using the lattice determination of the latter the authors  find
a very good agreement with the non-perturbative screening masses.


\section{Perturbative next-to-leading order}
\label{sec_nlo}
As we have anticipated, the recent technological developments
outlined in Sec.~\ref{chap_sumrule} make the extension to 
next-to-leading order of the
kinetic approach sketched in Sec.~\ref{chap_classical} possible.
In that Section we have introduced the three fundamental processes
that make up the leading-order collision operator, i.e. diffusion, large-angle scatterings
and collinear splittings/joinings. Next-to-leading order contributions represent a relative
$\OO(g)$ correction and come about because, as we have mentioned, soft gluons
are highly occupied, with $\nbe(q^0\sim gT)\approx T/q^0\sim 1/g$. For this reason
one can show that loops composed of soft propagators only are suppressed
by a factor of $g$ rather than $g^2$, giving rise to \emph{soft loop
corrections}. Another source of $\OO(g)$ corrections comes from regions of the
LO calculation where a gluon becomes soft, without being properly treated, i.e.
with HTL resummations. For instance, in large-angle scatterings, the
integrations over the energies  of the incoming and outgoing gluons stretch down to zero, including 
an $\OO(g)$ region of phase space where they are soft, but treated as massless, rather
than HTL, excitations. At NLO one then needs to replace the improper LO evaluation
of these \emph{mistreated regions} with the proper one. 

In the following we will proceed to sketch the kinematical regions
affected by NLO corrections and their evaluation. We will not dwell on mistreated
regions, which, in a nutshell, require the identification and subtraction
of the relevant counterterms in the corresponding points of the NLO calculations.

Large-angle scatterings are not sensitive to soft loop corrections, as they 
would require internal and/or external soft lines, which are either excluded
by construction or suppressed. Hard loops represent an $\OO(g^2)$ correction,
so that this region is not affected at NLO.

Collinear processes are sensitive to soft loop corrections: they give rise
to the $\OO(g)$ correction\cite{CaronHuot:2008ni} to $\cc(\xp)$ which we have 
mentioned in the previous Sections,
as well as to an $\OO(g)$ correction\cite{CaronHuot:2008uw}
to the dispersion relations of the hard, collinear particles.

Diffusion processes are also naturally sensitive to $\OO(g)$ corrections,
since the typical momenta there are by construction soft. The soft loop corrections
to $\qhat$\cite{CaronHuot:2008ni} have also been mentioned before,
while those to $\ql$\cite{jetwip} will be introduced in the following. 

Finally, a new process appears at NLO. We label it \emph{semi-collinear}. 
It is related to the example we made earlier on in this Section of a large-angle 
scattering with an incoming (or outgoing) soft gluon, which causes the angle
between the two outgoing (or incoming) hard ones to be smaller than in the 
large-angle case but larger than in the pure collinear case. Indeed,
as we shall see, this process will represent a bridge between 
the other three.

Schematically, the collision operator then takes this form
\begin{equation}
\delta C=\delta C_\mathrm{coll}+
\delta C_\mathrm{diff}[\mu]+
\delta C_\mathrm{semi-coll}[\mu],
\label{nlosketch}
\end{equation}
where we anticipate, as noted by the $\mu$ dependence,
that the last two processes will require regulation, but that the 
dependence on the regulator will vanish in the sum.

\subsection{The collinear region}
\label{sec_nlo_coll}

In the collinear regime the $\OO(g)$ corrections enter then in two places,
as we anticipated: both the effective thermal masses squared $\mmg$ and the collision kernel 
$\cc(\xp)$ get $\OO(g)$ corrections\footnote{\label{foot_counterterm}
For simplicity we do not illustate the two further mistreated
regions\cite{jetwip} in the soft and semi-collinear limits respectively.}.

For what concerns the thermal masses, an expression in terms
of gauge-invariant fermionic and bosonic condensates supported
on the light-cone is known\cite{CaronHuot:2008uw}. It 
reads\footnote{In QCD with quarks there is also a fermionic condensate. Both condensates
contribute to the asymptotic masses of gluons and quarks.}
\begin{equation}
	\mmg=g^2\ca\, Z_g\qquad Z_g \equiv\frac{1}{d_A}\left\langle
	v_\mu F^{\mu\nu a}\left(\frac{1}{(v\cdot D)^2}\right)_{ab}
	v_\rho F^{\rho\,b}_\nu\right\rangle,
	\label{asymmass}
\end{equation}
where $v=(1,\v)$ is a null vector ($\vert \v\vert =1$). Intuitively, $Z_g$ describes how thermal
fluctuations of the gauge fields affect the propagation of a fast,
light-like particle. It can be rewritten as
\begin{equation}
	Z_g=\frac{-1}{d_A}
\int_0^\infty dx^+ \, x^+ \langle \vphantom{\frac{1}{2}} v_{\mu}
F_a^{\mu\nu}(x^+) U^{ab}_{ A}(x^+,0)
 v_{\rho}{F_b^\rho}_\nu(0)\rangle,
	\label{zgwilson}
\end{equation}
which makes the connection with $\qhat$ clearer: by comparing
with \Eq{qT_versionb}, it is easy to see that they
share the same Lorentz structure. At leading order 
$Z_g$ is dominated by hard ($p\sim T$) momenta and one easily has
\begin{equation}
Z_g^{\rm LO} = 2\int \frac{d^3 p}{(2\pi)^3 p} \nbe(p) = \frac{T^2}{6}.
\label{Zg_LO}
\end{equation}
The $\OO(g)$ contribution arises from the proper treatment of the IR
limit of the above integration, requiring in principle HTL resummation.
However, $Z_g$ is supported on a lightcone, so that the methods of 
Sec.~\ref{chap_sumrule} become applicable. As shown by Caron-Huot\cite{CaronHuot:2008uw}
(further details can also be found here\cite{Ghiglieri:2013gia})
the $\OO(g)$ contribution is easily obtained from the Euclidean zero-modes, yielding
\begin{equation}
\delta Z_g = T\int\frac{d^3q}{(2\pi)^3} \frac{\qp^2}{(q_z+i\epsilon)^2}
\left( \frac{-1}{q^2+\md^2} + \frac{1}{q^2} \right)
 = - \frac{T\md}{2\pi},
\end{equation}
so that the NLO correction to the asymptotic mass $\delta \mmg$ reads
\begin{equation}
\label{minftyNLO}
\delta \mmg=g^2 C_{A}\left(-\frac{T \md}{2\pi}\right).
\end{equation}

\begin{figure}[ht]
\begin{center}
	\includegraphics[width=8cm]{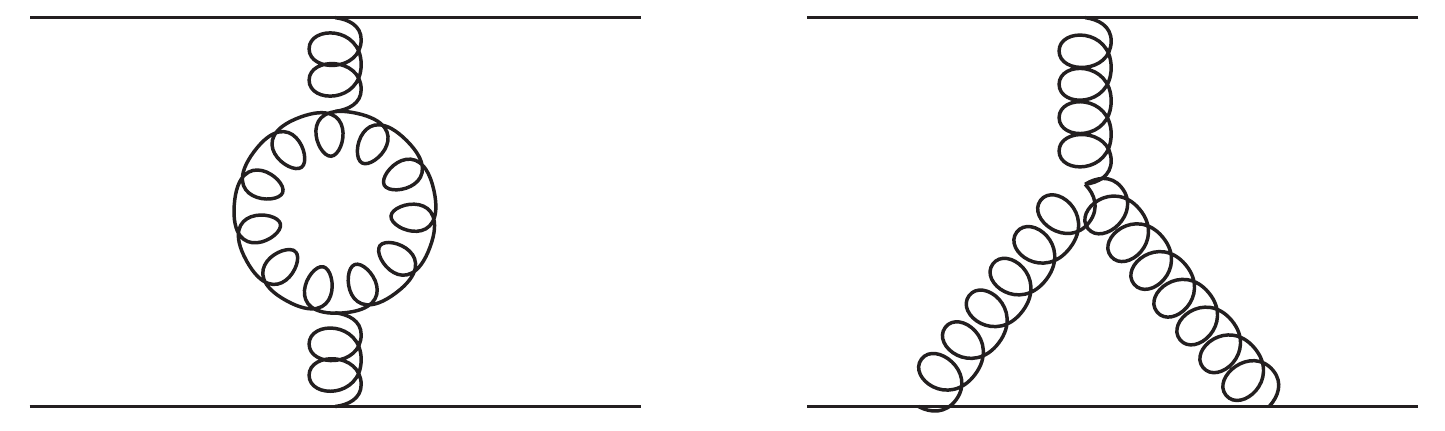}
\end{center}
\caption{Two sample diagrams necessary for the NLO evaluation of $\cc(\qp)$. The
	conventions are as in Fig.~\ref{fig_lo_cq}.}
\label{fig_nlo_cq}
\end{figure}

For what concerns the scattering kernel, Caron-Huot\cite{CaronHuot:2008ni}
has shown that three-body contributions vanish in the three-pole operator sketched
below Eq.~\eqref{bspace}. Hence, also at NLO one just needs the sum of three
two-body contributions at different transverse separations. The computation
of these is nothing but the NLO contribution to $\cc(\xp)$. 
We have shown in detail in Sec.~\ref{sub_sumrule} how to ``Euclideanize''
the leading-order calculation. At NLO one then needs to compute one-loop
diagrams in  three-dimensional EQCD, such as those shown in Fig.~\ref{fig_nlo_cq}.
The result for $\delta \cc(\qp)$\cite{CaronHuot:2008ni} and the corresponding
$\delta \cc(\xp)$\cite{Ghiglieri:2013gia} are not shown here for brevity. We
point to the references for their expressions.
We would like to stress again how the the reduction to the Euclidean, three-dimensional
theory makes this calculation tremendously simpler. If we were to compute
the diagrams of Fig.~\ref{fig_nlo_cq} in the HTL-resummed theory in four Minkowskian dimensions,
we would have to deal with complicated
denominators of (up to 4) HTL propagators (see Eq.~\eqref{htl})
and the intricate HTL vertices. Indeed, in both our example diagrams we would need to consider them along
the standard bare vertices. In the Euclideanized, three-dimensional theory, on the other hand, 
the propagators become simple massive or massless ones and the effective vertices 
just disappear, as they depend linearly on the external 
frequencies, which vanish in dimensional regularization.\footnote{Indeed, there are 
	no resummed vertices of the same order of the bare ones in EQCD, as we have shown
	in Sec.~\ref{chap_lattice}.}

In order to obtain the NLO correction to the collinear rate~\eqref{jmcoll},
Eq.~\eqref{bspace} is then to be perturbed by  $\delta \mmg$ and $\delta \cc$. The
first order in this perturbation $\delta \f(\b)$, is  plugged in Eqs.~\eqref{b_want} 
and \eqref{jmcoll}.
Details on this procedure may be found in these references\cite{Ghiglieri:2013gia,jetwip}.
Algorithms for the solution of the differential equation with the correct
boundary conditions are discussed in these other works\cite{Aurenche:2002wq,Ghiglieri:2013gia,Ghisoiu:2014mha,Ghiglieri:2014kma}.

We furthermore stress that, should a factorized approach based on lattice measurements
of soft observables, as outlined at the end of Sec.~\ref{chap_lattice}, be applied here,
one would need a non-perturbative measurement of the aforementioned three-pole operator.
It would be interesting to analyze the non-perturbative appearance of three-body forces
in such an object.

Finally, we remark that recently it has been
pointed out\cite{Liou:2013qya,Blaizot:2013vha} that collinear processes
are sensitive to a large, double-logarithmic correction, which
also shows up as a double-logarithmic correction to $\qhat$. This
term represent a relative $\OO(g^2)$ correction in our framework and
would hence contribute to NNLO. It would arise from considering
the first correction to the eikonal approximation in deriving
Eq.~\eqref{onetwocollision}.

\subsection{The diffusion  sector at NLO}
\label{sec_long_diff}
\begin{figure}[ht]
\begin{center}
\includegraphics[width=12.5cm]{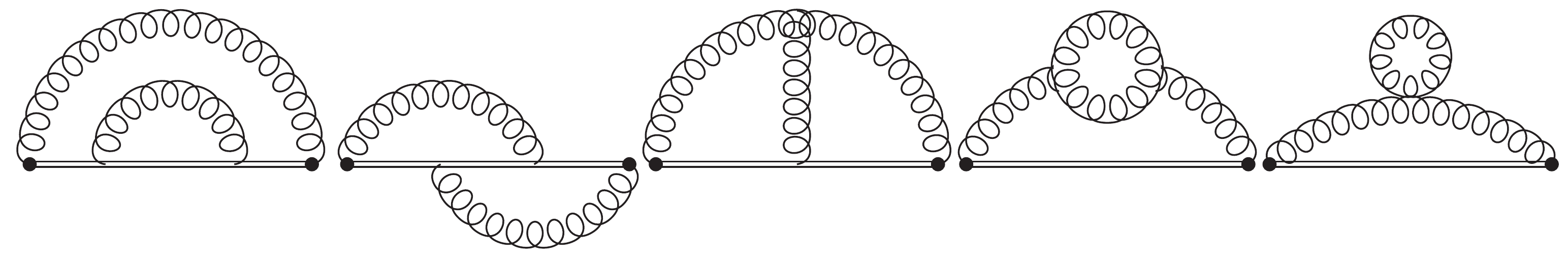}	
\end{center}
\caption{Diagrams contributing to $\delta\qhat$ and $\delta\ql$ at NLO. The blobs
represent again the field strength insertions. Three- and four-gluon vertices
are either bare or resummed HTL vertices.}
\label{fig_nlo_soft_diagrams}	
\end{figure}

As we anticipated, the Fokker-Planck diffusion equation~\eqref{fp} still applies.
We then need the $\OO(g)$ corrections to $\qhat$ and $\ql$, which
unambiguously determine $\hat{e}$ at NLO through the procedure sketched in 
Eq.~\eqref{einsteinrel}.  Both can be calculated 
from their Wilson line definitions~\eqref{qT} and \eqref{qL}. The former
is computed with the Euclidean technology of Sec.~\ref{sub_euclid} and 
the latter with the sum rules of Sec.~\ref{sub_sumrule}. However, before
going in the details of the two separate calculations, one can proceed
to a  
simplification \cite{simonguy,jetwip}, which is attained by using the equation 
of motion of the Wilson line  $D_{x^+}^-U(x^+;0)=0$ to simplify
$F^{-\mu}$ to $\partial^\mu A^-$ (in all gauges but the $A^-=0$ one) and
by noting again that operator ordering is not relevant at NLO: classical
soft gluons 
commute, as we have used in obtaining the simplified
forms~\eqref{qT_versionb} and \eqref{qL_versionb}.
The Wilson line structure  then simplifies to
\begin{eqnarray}
\label{qhatsimon}
\qhat&=&\frac{g^2 \crr}{d_A}\int_{-\infty}^{+\infty}dx^+\left\langle  \partial^\perp
A^{-\,a}(x^+)U_A^{ab}(x^+,0)\partial^{\perp}A^{-\,b}(0)\right\rangle,\\
\label{defqlongsimon2}
\ql&=&\frac{g^2 \crr}{d_A}\int_{-\infty}^{+\infty}dx^+\left\langle  \partial^+
A^{-\,a}(x^+)U_A^{ab}(x^+,0)\partial^{+}A^{-\,b}(0)\right\rangle.
\end{eqnarray}
Their evaluation
requires the computation of the diagrams shown in Fig.~\ref{fig_nlo_soft_diagrams}.

For what concerns $\qhat$, the reduction to the Euclidean, three-dimensional
theory introduced before makes this calculation too tremendously 
simpler. 
One then obtains for $\delta \qhat$, the NLO contribution to $\qhat$ \cite{CaronHuot:2008ni}
\footnote{Caron-Huot obtained this result by applying Eq.~\eqref{qhat_LO} to his 
calculation of $\cc(\qp)$ at NLO. The two methods are completely equivalent,
as Eq.~\eqref{qT} can be formally derived from Eqs.~\eqref{defcq} and \eqref{qhat_LO}\
when $\cc(\qp)$
is defined from the Wilson loop in Fig.~\ref{wilsonfig}, as shown in this reference\cite{Benzke:2012sz}.}
\begin{equation}
\label{nloqhat}
\delta \qhat=\frac{g^4\crr\ca T^3}{32\pi^2}\frac{\md}{T}\left(3\pi^2+10-4\ln 2\right).
\end{equation}

For what concerns  $\delta \ql$, the sum-rule technology we have introduced
in Sec.~\ref{sub_sumrule} can be applied and leads to another tremendous 
computational simplification\cite{jetwip}. As a general strategy, we remark
that the Wilson line
propagators depend only on the minus components of the momenta, so that
 the plus component, which we call $\qll$, can be deformed again.
This corresponds to expanding those diagrams for large, complex $\qll$.
The leading contribution can be of order $(\qll)^0$ and the subleading one of order $(\qll)^{-1}$.
Higher-order terms are suppressed  and can be neglected. The leading, $\OO((\qll)^0)$ term, once integrated
along the contour, will give rise to a linear divergence, which has to match with 
the one arising from the corresponding mistreated region in the collinear process
(see footnote~\ref{foot_counterterm}). 
Indeed, the calculation\cite{jetwip} confirms that the two divergences cancel.

For what concerns the $\OO(1/\qll)$ term, we already encountered such behavior at LO, where it gave
rise to the asymptotic mass. In the photon case \cite{Ghiglieri:2013gia} it was found,
rather surprisingly, that the NLO contribution to this term
amounted to replacing the quark asymptotic mass $\mmf$ with
$\mmf+\delta\mmf$, where $\delta \mmf$ is its soft $\OO(g)$ correction, in the
fermionic equivalent of Eq.~\eqref{lofinal} and then expanding to linear order in $\delta \mmf$
(and hence in $g$). The explicit sum-rule computation\cite{jetwip} of the diagrams 
in Fig.~\ref{fig_nlo_soft_diagrams} yields the same (up to the different asymptotic mass)
i.e.
\begin{equation}
\label{soft_guess}
\frac{\mmg+\delta\mmg}{\qp^2+\mmg+\delta \mmg}
=\frac{\mmg}{\qp^2+\mmg}+\delta\mmg\frac{\qp^2}{(\qp^2+\mmg)^2}+\OO(g^2)\,,
\end{equation}
where $\delta \mmg$ is given in Eq.~\eqref{minftyNLO}.
This result can be interpreted physically in the following way: expanding for large (complex)
$q^+$ takes the soft fields to approach their collinear limit, where the only
effect of resummation is to introduce an asymptotic mass (at leading order) and the soft 
correction thereto (at next-to-leading order). For this same reason, 
HTL vertices become small and not relevant.

Hence we obtain
\begin{equation}
\delta\ql
=g^2\crr  T
\int\frac{d^2\qp}{(2\pi)^2}
\frac{\qp^2\delta\mmg}{(\qp^2+\mmg)^2}=\frac{g^2\crr  T\delta\mmg}{4\pi}\left[\ln
\left(\frac{\left(\mu^\mathrm{NLO}\right)^2}{\mmg}\right)-1\right],
\label{finallongdiffnlo}
\end{equation}
where we have introduced a regulator $\mu^\NLO$. As we will show, the semi-collinear region
will remove the dependence on it, so that it should be taken to obey $gT\ll\mu^\NLO\ll\sqrt{g}T$.

\subsection{The semi-collinear region}
\label{sec_semi}
\begin{figure}[ht]
\begin{center}
\includegraphics[width=13cm]{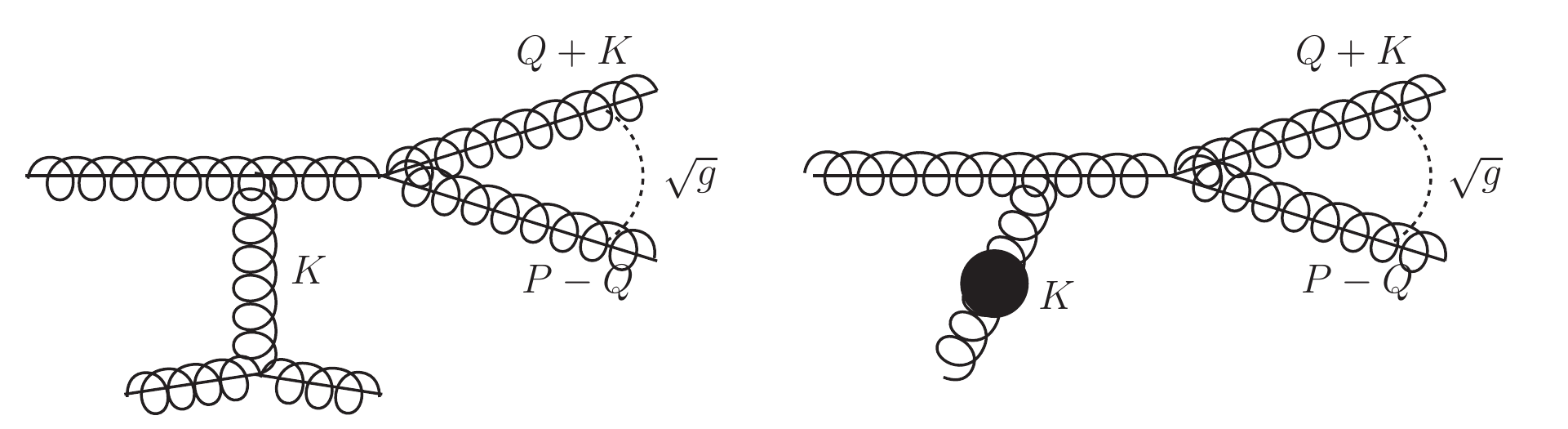}	
\end{center}
\caption{Diagrams for two typical semi-collinear processes. In the first case the soft gluon
is in the spacelike Landau cut, whereas in the second case it is on its timelike plasmon pole, 
represented by the black blob.}
\label{fig_semicoll}
\end{figure}

As we anticipated before, semi-collinear processes
can be seen as $\onetwo$ splitting processes where the opening angle
(and hence the virtuality) are larger. Two examples are drawn in Fig.~\ref{fig_semicoll}.
The scalings of this region are as follows: $K\sim gT$ is soft, whereas the two final-state particles
are collinear, albeit with an increased virtuality and opening angle with respect to the collinear sector.
The leading contribution then comes from $\qll\sim T,\,\qm\sim gT,\,\qp^2\sim gT^2$, $Q^2\sim gT^2$.

Naive power-counting arguments would suggest
that the semi-collinear
 region should contribute to leading order, as it is the largest slice of phase space 
where a soft gluon can attach to a $\onetwo$ process. However, once all diagrams are summed
and squared, a cancellation, first noticed  in the context of photon
radiation\cite{Arnold:2001ba}, introduces an extra $\OO(g)$ suppression. Furthermore,
since $K\sim gT$ in all components, the contribution from
timelike soft gluons, e.g. plasmons, is now allowed. This is contrasted
by the collinear region, where kinematics enforce $k^-\sim \delta E\sim g^2T\ll k^+,\kp$,
thus restricting soft gluons to the space-like domain only.

The contribution $\delta C_\mathrm{semi-coll}$ to the collision operator can be written 
in the same way as the collinear one, as given by Eq.~\eqref{onetwocollision},
with the replacement of the collinear rates with semi-collinear ones.
In pure glue it reads \cite{jetwip} 
\begin{eqnarray}
	\nn \delta C_\mathrm{semi-coll}[\mu]&=&\int_{-\infty}^{+\infty}d\omega\left[ 
		f_{p+\omega}\frac{d\Gamma(p+\omega,\omega)}{d\omega}
\bigg\vert_\mathrm{semi-coll}\right.\\
&&\left.\hspace{2cm}-f_p
\theta(p-2\omega)
\frac{d\Gamma(p,\omega)}{d\omega}\bigg\vert_\mathrm{semi-coll}\right],
\label{semicollop}
\end{eqnarray}
The derivation of the semi-collinear rates then
requires the evaluation of processes of the form of Fig.~\ref{fig_semicoll}, with $p,\qll\gg
\qp\gg\kp,\kl$. This results in the following factorized form\cite{jetwip}
\begin{equation}
	\frac{d\Gamma(p,\omega)}{d\omega}\bigg\vert
_\mathrm{semi-coll}=\frac{g^2\ca \, }{\pi p}\frac{(1-x+x^2)^3}{x(1-x)}(1+ n(\omega))
(1+ n(p-\omega))\int\frac{d^2\qp}{(2\pi)^2}
\frac{\qhat(\delta E)}{ \qp^4}
\label{colltosemi}
\end{equation}
where 
\begin{equation}
\delta E = \frac{p\,\qp^2}{2\omega(p-\omega)},
\label{deltaesc}
\end{equation}
and $\qhat(\delta E)$, which was first introduced for photon radiation 
\cite{Ghiglieri:2013gia}, is a modified version of $\qhat$ which
keeps track of the changes in the small light-cone components $p^-$ and
$q^-$, induced by the interaction with the soft degrees of freedom.

In the field-theoretical language of Sec.~\ref{chap_classical},
$\qhat(\delta E)$ is given by\footnote{Strictly speaking,
	this is the version that applies to photon radiation and 
	in this case we would need a more complicated three-pole expression,
	along the lines of the discussion that followed Eq.~\eqref{bspace}. 
	However, we are evaluating it at the leading-order, one-gluon 
exchange level, so that the more complicated form is not necessary.}
\begin{equation}
\label{Isc}
\qhat(\delta E)=\frac{g^2\crr}{d_A} \int_{-\infty}^\infty dx^+\,e^{ix^+\delta E} \,
 \langle v^\mu {F_\mu}^{\nu, a}(x^+)
 U^{ab}_{A}(x^+,0) 
v^\rho F^b_{\rho\nu}(0)\rangle,
\end{equation}
where we have made use of the simplification discussed in Eq.~\eqref{qhatsimon}.
The dependence on the minus components appears as a simple phase accumulation
during the eikonal propagation, so that one can also easily see how
Eq.~\eqref{Isc}  goes into Eq.~\eqref{qT} for $\delta E\to 0$. 
Eq.~\eqref{Isc} can also be evaluated using Euclidean techniques, yielding \cite{Ghiglieri:2013gia}
\begin{equation}
\label{semiintexp}
\qhat(\delta E)=g^2\crr T\int\frac{d^2\kp}{(2\pi)^2}\bigg[\frac{\md^2\kp^2}
{(\kp^2+\delta E^2)(\kp^2+\delta E^2+\md^2)}+\frac{2\delta E^2}{\kp^2+\delta E^2}\bigg].
\end{equation}

As we have anticipated at the beginning of this Section,
a counterterm from the large-angle scattering region needs to be subtracted, as well as a
collinear one (see footnote~\ref{foot_counterterm}). The former arises from taking the $\md\to 0$ limit of Eq.~\eqref{semiintexp}
(there is no resummation taking place on external legs at LO) and the latter by taking the 
$\delta E\to 0$ limit (changes in the minus component are not accounted for in the
collinear limit). Once these subtractions are performed, the semi-collinear rate becomes
\begin{eqnarray}
\nn	\frac{d\Gamma(p,\omega)}{d\omega}\bigg\vert_\mathrm{semi-coll}&=&\frac{g^4\ca^2 T }{\pi p}
(1+ n(\omega))(1+ n(p-\omega))\int\frac{d^2\qp}{(2\pi)^2}
\int\frac{d^2\kp}{(2\pi)^2}\frac{1}{\qp^4}\\
\nn&&\times
\frac{(1-x+x^2)^3}{x(1-x)}
\bigg[\frac{\md^2\kp^2}
{(\kp^2+\delta E^2)(\kp^2+\delta E^2+\md^2)}-\frac{\md^2}{\kp^2+\md^2}\bigg].\\
&&
\label{jmsemicoll}
\end{eqnarray}
We stress that the  semi-collinear collision operator is given by 
Eq.~\eqref{semicollop}.

The $\qp$ integration in Eq.~\eqref{jmsemicoll} is to be understood as IR-regulated by $\mu^\NLO$.
One can show\cite{jetwip} how the small-$\omega$-and-$\qp$ region gives rise to IR logarithms
that cancel the $\mu^\NLO$ dependence of diffusion processes. Details of how 
the transverse integrations can be carried out analytically are also given there. The $\omega$
integration remains to be performed numerically.

\section{Summary and conclusions}
\label{sec_concl}
In this review we have shown how at leading and next-to-leading order the propagation, energy loss,
and momentum diffusion of high-energy particles in the QGP can be
cast in the form of a Boltzmann equation 
describing
the interactions between the jet particles and the hard and soft constituents
of the plasma (\Eq{loboltzmann} and \Eq{nlosketch}). In \sect{chap_classical}
we introduced this formalism at leading order, where three processes
contribute. These processes include   $\twotwo$ scatterings of the jet with the hard 
particles,  drag and (longitudinal and transverse) momentum diffusion 
induced by the soft highly occupied gluonic background, 
and collinear bremsstrahlung (and merging) induced by the transverse momentum diffusion. The $\twotwo$
scatterings are dealt with in a standard kinetic fashion 
(\Eq{eq:collision22exp}), the collinear splittings
and joinings give rise to a somewhat simpler rate term (\Eq{onetwocollision}),
and the interactions with the soft background are described by a Fokker-Planck
equation (\Eq{fp}) or equivalently by a Langevin equation (\Eq{langevin}).

We have furthermore shown how the contributions of the soft degrees 
of freedom to the kinetic picture (the diffusion coefficients in the
Fokker-Planck equation and the transverse scattering kernel in the collinear rate)
can be cleanly defined as correlators of Wilson lines, or Wilson loops,
supported on the lightcone or on a light front (see Eqs.~\eqref{qT}, \eqref{qL}
and \eqref{cfromw}). These Wilson lines are to be evaluated using the Hard Thermal
Loop effective theory. We devote \sect{chap_sumrule} to the illustration
of the recent understanding of thermal field theory (including Hard
Thermal Loops) on the lightcone. We show how the causal properties of 
amplitudes at light-like separations bring about a dramatic simplification
in their calculation\cite{CaronHuot:2008ni}. To put it simply,
jet and hard particles, which
move at $v\sim c$, probe the medium in a very simple (thermodynamical) way,
since the
soft background ``does not have the time'' to respond to the perturbation
they cause. At a practical level, we show how the computation of certain lightlike correlators, such as transverse momentum diffusion, ``euclideanizes'', that is, it reduces
to a much simpler calculation in the three-dimensional Euclidean theory EQCD
(see Eqs.~\eqref{simonmagic} and \eqref{C_LO}). We also
show how a different set of lightlike correlators, such as
 longitudinal diffusion, does not euclideanize. This case also simplifies, 
however, and the same causal properties on the lightcone make it sensitive only to the thermal mass and soft corrections thereto  
(see Eqs.~\eqref{arcexpand} and \eqref{lofinal}).

In Sec.~\ref{chap_lattice} and Sec.~\ref{sec_nlo} we reviewed two major consequences of these lightcone simplifications. First, we showed how operators that euclideanize
can be measured on the lattice. Second, we reviewed the consequences on 
perturbation theory. At leading order the lightcone simplifications result in simple analytical closed forms in place of relatively straightforward numerical
integrals. At NLO the impact is much more far-reaching -- calculations that
would have been extremely intricate, if not impossible, can now be
performed with relative ease.

In more detail, in \sect{chap_lattice} we reviewed the basics of EQCD, and 
described the first non-perturbative calculations of  
$\qhat$ and the transverse
momentum scattering kernel using this effective theory~\cite{Panero:2013pla}. We would like to emphasize how these first
results open a new avenue of research -- all other Euclidean operators can 
be measured on the lattice in the same way. This  creates the tantalizing possibility
of a factorized approach to kinetics,  where perturbation theory is used at the
hard scale to compute the $\twotwo$ scatterings and the splitting processes, while
the 3D lattice is used at the soft and ultrasoft scales to compute the diffusion processes and  the transverse scatterings leading to bremsstrahlung.

In \sect{sec_nlo} we give an overview of the NLO generalization 
of the LO Boltzmann equation. We showed  how the three
processes at LO are corrected at NLO (see \Eq{nlosketch}). 
Besides the corrections  to the LO diffusion and collinear processes,
a new \emph{semi-collinear} process enters, which appears as a bridge between
the others, being collinear but with relaxed constraints, i.e. going beyond
strict collinearity. In the evaluation of each of these processes we sketched
how the lightcone simplifications described above make the NLO analysis possible.
We described the NLO calculations of 
transverse\cite{CaronHuot:2008ni} and longitudinal\cite{jetwip} momentum diffusion (\Eq{nloqhat} and \Eq{finallongdiffnlo}), 
and introduced two related Euclidean operators: the
condensate $Z_g$ responsible for the thermal mass (\Eq{asymmass}), and
$\qhat(\delta E)$, a modified version of $\qhat$ relevant for the semicollinear
bremsstrahlung (\Eq{Isc}).  The corrections to the longitudinal diffusion and
drag
coefficients at NLO are due to corrections to  the thermal mass (or $Z_g$).
Finally, we presented complete expressions 
for the NLO rates in the Boltzmann equation for pure gauge theory.

The extension of the NLO Boltzmann equation to
include light quarks 
requires one extra process at leading- and next-to-leading order,
the \emph{conversion process}, which can be seen as a fermionic analogue of diffusion~\cite{jetwip}.
The interaction of hard or jet particles with soft quarks results
in a conversion from a quark to a gluon with approximately the same momentum
and vice-versa.  
The computation of these rates involve fermionic correlators on the light cone.
It turns out that the lightcone simplifications of the fermionic HTLs are strikingly similar 
to the longitudinal diffusion described in Sec.~\ref{sub_sumrule}, and consequently
the conversion rates are sensitive only to  the (fermionic) asymptotic mass and
its soft corrections. These corrections to the fermion mass are computable with the 3D effective theory, in much the way that the effective theory can be used to compute corrections to $Z_g$.

One very important question that we leave to future works is the impact
of these developments on calculations of jet energy loss and their comparison
to experimental data. We remark that the Monte Carlo event generator
MARTINI\cite{Schenke:2009gb} currently implements a kinetic approach
that is very close to the leading-order picture sketched in 
\sect{chap_classical}. It is then an ideal candidate for the inclusion
of the developments discussed in \sect{chap_lattice} and \sect{sec_nlo}.
The reorganization in terms of $\twotwo$ scattering, diffusion and collinear processes,
as well as the implementation of the NLO corrections discussed in
\sect{sec_nlo}, is already underway. This can be easily complemented
by the inclusion of non-perturbative input. Besides the measured
transverse scattering kernel, lattice calculations for $\qhat(\delta E)$ and
$\mmg$ could easily be included into MARTINI, should they become available in the future.
It would also be interesting to see the results of this numerical 
implementation for the angular structure of jets,
for which recent order-of-magnitude estimates from perturbation theory
are available\cite{Kurkela:2014tla}.

\begin{figure}[ht]
\begin{center}
	\includegraphics[width=8cm]{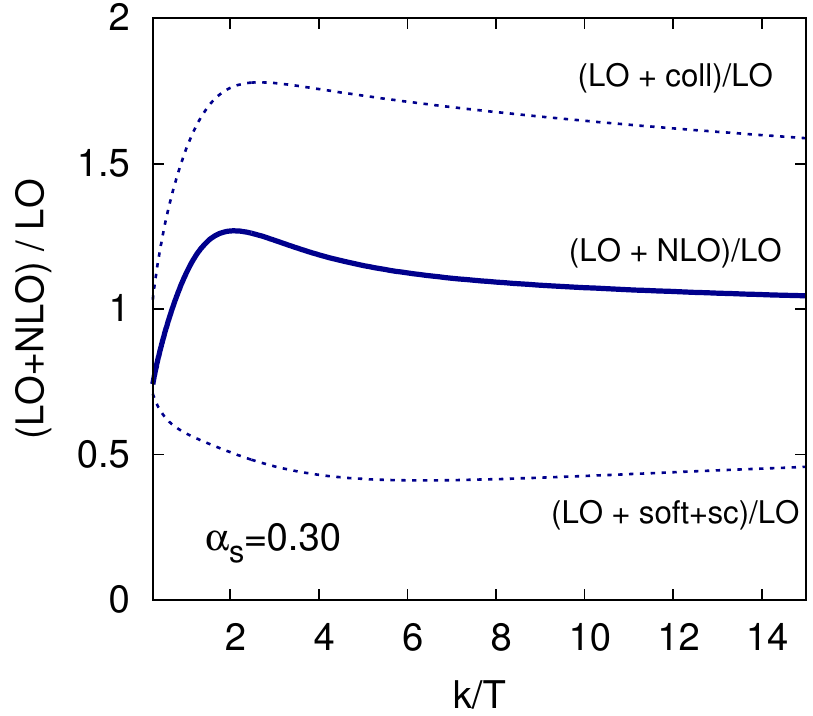}
\end{center}
\caption{The differential photon emission
    rate $d\Gamma_\gamma/dk$ relative 
to the leading order rate as a function of the photon momentum $k/T$
at a fixed coupling $\als=0.3$ in a $N_{f}=3$ QGP.
The full next-to-leading order rate\cite{Ghiglieri:2013gia} (LO+NLO) is a sum of the
leading order rate (LO), a collinear correction (coll), and 
a soft+semi-collinear correction (soft+sc).
The dashed curve labeled LO+coll shows the ratio of rates when 
only the collinear correction is
included, with the analogous notation for the 
LO + soft+sc curve. 
Figure taken from the original reference\cite{Ghiglieri:2013gia}. }
\label{fig_nlo_photon}
\end{figure}

As a general remark, we find it difficult to tell \emph{a priori}
the impact the NLO corrections of \sect{sec_nlo} will have. The
recent calculation to NLO of the thermal photon rate\cite{Ghiglieri:2013gia},
(which includes
many of the ingredients presented here, such as Euclideanizations, 
light-cone causality, collinear and semi-collinear processes) showed
how the NLO corrections  naturally divide into two large, and 
largely canceling, contributions -- see Fig.~\ref{fig_nlo_photon}. 
Clearly the NLO corrections are modest over a significant range of 
photon momentum.
The positive contribution arose from collinear
processes enhanced by a NLO increase of the scattering kernel, and
a NLO decrease of the thermal mass, while the negative contribution arose 
from the semi-collinear and NLO soft processes.
The large cancellation between these two corrections 
is largely accidental and is dependent on the details of the medium,
such as the number of colors and  flavors. Nevertheless, we anticipate a similar
cancellation for the current more complicated case of parton energy loss.

Finally, we note that the kinetic approach we have outlined is often
used to compute the  transport 
coefficients of QCD plasmas~\cite{Arnold:2000dr,Arnold:2003zc}.
While an extension to NLO of these calculations 
requires more than what is outlined in \sect{sec_nlo},
 a computation of transport coefficients in thermal QCD beyond leading order will almost certainly make considerable use of the euclidean light-cone simplifications described 
in this review.

\section*{Acknowledgments}
We would like to thank Guy Moore for collaboration in this reference\cite{jetwip}.
We also thank Simon Caron-Huot, Aleksi Kurkela, Mikko Laine and Marco Panero for useful
conversations.
This work was supported
in part by
the Swiss National Science Foundation
(SNF) under grant 200020\_155935 and a grant from the U.S. Department of Energy, DE-FG-02-08ER4154.
\bibliographystyle{ws-rv-van}
\bibliography{eloss.bib}

\printindex                         
\end{document}